# Vibrational and nonlinear optical properties of amine-capped push-pull polyynes by infrared and Raman spectroscopy

*Pietro Marabotti†, Alberto Milani†, Andrea Lucotti‡, Luigi Brambilla‡, Matteo Tommasini‡, Chiara Castiglioni‡, Patrycja Męcik§, Bartłomiej Pigulski§, Sławomir Szafert§[\*], Carlo Spartaco Casari†[\*]*

† Micro and Nanostructured Materials Laboratory - NanoLab, Department of Energy, Politecnico di Milano via Ponzio 34/3, I-20133, Milano, Italy
‡ Department of Chemistry, Materials and Chem. Eng. 'G. Natta', Politecnico di Milano Piazza Leonardo da Vinci 32, I-20133, Milano, Italy
§ Faculty of Chemistry, University of Wrocław, 14 F. Joliot-Curie, 50-383 Wrocław, Poland

## Abstract

The vibrational properties of a series of push-pull polyynes have been studied by infrared and Raman spectroscopy. The simultaneous activation of a strong infrared and Raman mode, i.e. the collective vibration of CC bonds of the sp carbon chain, highlights the effectiveness of a polyyne bridge in promoting charge transfer between the electron donor (D) and acceptor (A) ends, thus suggesting that *ad-hoc* functionalized polyynes are worth being explored as building blocks of organic materials with attractive first-order optical nonlinearity. The optical, electronic, and vibrational properties of these molecules have been investigated with the support of density functional theory calculations, as well as the electronic and vibrational first hyperpolarizabilities (β). The mid-low region of the IR spectra (800-1600 cm$^{-1}$) has been analyzed in detail, searching for marker bands of the specific terminations of the different sp carbon chains thus achieving a complete vibrational characterization of sp hybridized push-pull systems.

## 1. Introduction

Polyynes are π-conjugated systems formed by sp-hybridized carbon chains featuring alternated triple and single carbon bonds[1–5]. As a model of finite carbon atomic wires representing carbyne, the interest in polyynes has grown during the last decades thanks to their predicted optical, electronic, mechanical properties and their possible applications in nanotechnology fields[6–8]. In particular, polyynes possess appealing nonlinear optical (NLO) properties as already observed in

[\*]*Corresponding authors: Tel. +39 02 2399 6331* Email: carlo.casari@polimi.it
Tel. +48 71 375 71 22 Email: slawomir.szafert@chem.uni.wroc.pl

other conjugated materials[2,9–12]. Moreover, the possibility to easily simulate one-dimensional models attracted the interest of theoreticians[13–19], providing great support in the analysis of vibrational experimental data of polyynes[13,20–25].

Polyynes can be synthesized by chemical and physical methods, such as arc discharge in liquid and pulsed laser ablation in liquid[1,4,5,24,26]. During the last years, the research on the chemical synthesis of polyynes has made significant steps, concerning the control of their chemical structure and their precise length[4,26,27]. One critical issue in the study of sp-hybridized carbon compounds is their instability under ambient conditions[5,28]. Indeed, carbon atomic wires in condensed phases tend to rearrange into more stable $sp^2$ systems, through crosslinking or oxidation reactions[28]. However, it has been proven that the presence of bulky endgroups hinders crosslinking, increasing the stability of the molecules up to the remarkable length of 48 carbon atoms[26,29,30]. Moreover, it was demonstrated that the encapsulation of polyynes in rotaxanes or carbon nanotubes allowed to synthesize and stabilize long linear carbon chains [31–34].

The successful strategies introduced for improving polyynes stability have allowed material scientists to investigate their properties in deeper detail. Regardless of their specific chemical structure (i.e., endgroups), all carbon atomic wires feature an intense Raman mode which behavior has been analyzed in the framework of the effective conjugation coordinate (ECC) theory applied to the π-conjugated polyyne [23,35,36]. Such strong Raman transition is assigned to the collective vibration of all the CC bonds of the sp chain and it is located in a spectral region where the other carbon allotropes are Raman inactive, which makes it a distinct fingerprint of these compounds[20,37]. The ECC band can give information about the structure, the length, and the terminations of the chains, and several other properties related to the presence of delocalized π electrons[14,20,21]. The occurrence of bending of the chain structure can be revealed through the comparative study of the ECC modes in IR and Raman spectra[22–24]. Indeed, the symmetry lowering determined by the deviation from a linear polyyne chain, as indeed it happens both in solid crystalline and solution phases, results in the activation of several marker bands which can be easily recognized[22–24].

Following the early insight that molecules with delocalized and polarizable π electrons were promising candidates for application in the field of optics and optoelectronics, thanks to their NLO performances, great efforts were spent on the design and synthesis of new organic molecules with enhanced hyperpolarizabilities and the rationalization of their NLO properties through theoretical models[38–50]. Most of the NLO effects of interest for technological applications occur with visible or near-infrared radiation so that the *electronic* hyperpolarizabilities are the quantities of interest for the design and optimization of the devices. However, the *vibrational* contribution to the NLO response of organic molecules is often far from being negligible[51–55]. A clear correlation between the electronic and vibrational hyperpolarizabilities was observed and ascribed to the strong electron-phonon coupling in molecules with conjugated π electrons[51,55]. For this reason, the vibrational hyperpolarizabilities can provide a good estimate of their electronic counterpart[56]. The quantitative agreement between the vibrational and the electronic hyperpolarizabilities was observed to improve with increasing hyperpolarizability values [51,55,57]. In this context, polyynes featuring two different molecular endgroups represent an ideal candidate to investigate vibrational and NLO properties with particular attention to their first hyperpolarizability that can be derived from vibrational spectroscopy investigations. The two different endgroups can induce the polarization of the sp chain that results in the activation in the IR of the characteristic Raman-active ECC modes of the sp chain[22]. Hence, remarkable spectroscopic features are expected if the end groups can act as donor (D) and acceptor (A) units, thus inducing a large static dipole moment along the polyyne axis, as observed in the case of push-pull polyenes and other D-π-A systems[38,45,46,58]. Unfortunately, the experimental studies of the NLO behavior of polyynes are limited by the availability of chemically stable samples of structurally well-defined compounds. The available experimental determinations deal with the electronic[9,10,59] and the vibrational[11,60] second hyperpolarizability (γ), while no experimental studies on push-pull polyynes showing large first hyperpolarizability (β) have been appeared so far, except for quadrupolar molecules with polyyne bridges, for which the first order NLO response has been measured by two-photon absorption experiments[51].

In this work, we show how in these push-pull systems the peculiar IR activity of the ECC modes is responsible for a large vibrational contribution to the first molecular hyperpolarizability. We also highlight the close correlation existing between the vibrational and the electronic contribution to the first molecular hyperpolarizability of push-pull polyynes. Such correlation was demonstrated in the past for push-pull polyenes and other organic molecules characterized by a backbone of sp$^2$ carbon atoms sharing a polarizable π electrons system[45,51,59,61]. The NLO response of push-pull polyynes and apolar (symmetric-ends capped) polyynes has been investigated by quantum chemical calculations of the electronic and vibrational hyperpolarizabilities of series of model molecules[57,62–65]. The effect of increasing the chain length, and the performances of polyene *vs* polyyne π bridges, as well as the relationship between electronic and vibrational hyperpolarizabilities have been analyzed in detail[57,62]. We investigate the IR and Raman spectroscopy of a series of stable push-pull polyynes that have been named ynamines due to their amine capping (see Fig. 1). We examine their spectroscopic behavior to gather experimental evidence that supports the remarkable β values determined by quantum chemical approaches. We have restrained ourselves to experimentally determine the $β_v$ values through the measurement of the vibrational intensities, by the procedure described in the work of Castiglioni et al.[56], because the strong dipole of ynamines is likely to drive aggregation effects in the solution state. Of course, such aggregation effects may significantly influence the assessment of the absolute IR and Raman intensities, and we plan to address this delicate issue in dedicated work.

The ynamines considered in this work have a varying chain length from 2 to 4 triple bonds and are terminated by a common acceptor at one end (A = benzonitrile) and by a varying donor (D) group at the other end. The analysis of their IR and Raman spectra shows the presence of the ECC transition as a strong Raman *and* IR band. This behavior is accounted for by DFT calculations, and it is justified by the sizeable charge transfer established between the two polar endgroups of ynamines. This electron charge transfer from the donor to the acceptor endgroup is sustained by the π electron cloud and determines a geometry relaxation of the whole polyyne chain. Remarkably, the CC bonds belonging to the chain become polarized. Hence, the CC stretching vibrations that are usually IR weakly active or inactive gain large infrared intensity, as it is was discussed for the ECC CC stretching vibrations of push-pull polyenes[45,46,56]. In push-pull polyenes, the simultaneous IR and Raman activity of the ECC mode, which originates the most intense band both in the IR and Raman spectra, provides the experimental signature of a large vibrational β. To the best of our knowledge, the present study is the first detailed characterization of the vibrational structure of a representative series of push-pull polyynes (ynamines) that highlights behavior that parallels one of the push-pull polyenes.

Besides the activation in the IR of the ECC modes, ynamines also show the characteristic evolution of structural, optical, and vibrational properties as a function of increasing π-electrons conjugation (*i.e.*, chain length)[5,21]. The bond length alternation parameter (BLA, i.e. the difference between the averaged single and triple bond lengths), the HOMO-LUMO gap, and the position of the ECC peaks, all depend on the sp chain length, as well as on the occurrence of different donor groups in the ynamines series considered. Finally, the 800 – 1600 cm$^{-1}$ region of the IR spectra of ynamines has been analyzed to assign the main markers of the different donor groups.

## 2. Materials and Methods

Figure 1 shows the chemical structures of the ynamines here investigated, grouped by their donor (D) group (the acceptor (A) group is the same for all the compounds). Powder samples exhibit chromatic tones going from yellow to red. Ynamines belonging to the A, B, and C series were synthesized through the method based on the one described in the work of Pigulski *et al.*[29]. The description of the synthesis of the molecules belonging to the D[n] series follows here below. For spectroscopic characterization of new compounds please see Supporting Information.

**D[2]:** 4-(Bromobutadiynyl)benzonitrile[66] (0.140 g, 0.608 mmol) was dissolved in dry MeCN (10 mL) and placed in a 20 mL screw-sealed vial. Next Na$_3$PO$_4$ (0.207 g, 1.22 mmol) and morpholine (0.080 mL, 0.91 mmol) were added. The mixture was stirred at 40 °C for 1 h and after that time 10 mL of diethyl ether was added. The reaction mixture was passed through an alumina plug (basic, Brockmann grade I, Et$_2$O) and solvents were evaporated to produce D[2] as a yellow solid (0.112 g, 0.474 mmol). Yield: 78%.

**D[3]:** 4-(Bromohexatriynyl)benzonitrile[66] (0.095 g, 0.315 mmol), dry MeCN (10 mL), Na$_3$PO$_4$ (0.107 g, 0.629 mmol), and morpholine (0.041 mL, 0.47 mmol) were reacted as for D[2]. Analogous workup gave orange solid of D[3] (0.053 g, 0.204 mmol). Yield: 65%.

**D[4]:** 4-(Iodooctatetraynyl)benzonitrile[67] (0.089 g, 0.277 mmol), dry MeCN (10 mL), Na$_3$PO$_4$ (0.091 g, 0.554 mmol), and morpholine (0.036 mL, 0.41 mmol) were reacted as for D[2]. Analogous workup gave D[4] as yellow solid (0.045 g, 0.158 mmol). Yield: 57%.

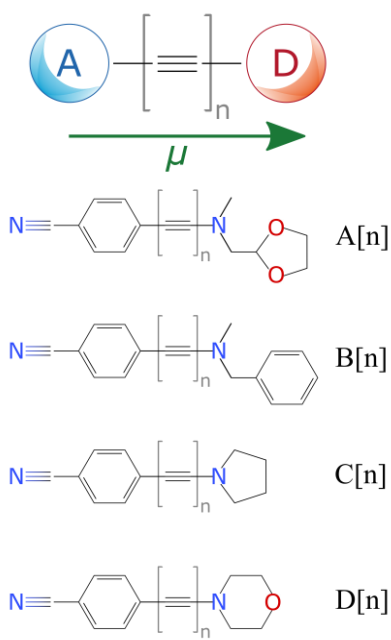

*Figure 1. Schematic representation and molecular structure of ynamines. [n] indicates the number of triple bonds in the sp chain. The green arrow represents the expected direction of the electric dipole moment.*

FT-Raman and IR spectra were recorded both in solid-state and in chloroform solution (chloroform, 99.8+%, stabilized with amylene, Fisher Chemical). The concentration of each ynamine in the corresponding solution was approximately $10^{-2}$ M. FT-Raman measurements were performed at room temperature with a Nicolet NXR9650 spectrometer (4 cm$^{-1}$ resolution, 50 µm spot size) equipped with a Nd-YVO$_4$ solid-state laser providing a 1064 nm excitation line. 512 scans were accumulated for each FT-Raman spectrum to reach an acceptable signal-to-noise ratio. Powders were probed at a power of approximately 400 mW while solutions at approximately 1 W. IR measurements were carried out with a Nicolet Nexus FTIR spectrometer coupled with a Thermo-Nicolet Continuµm infrared microscope and liquid nitrogen cooled MCT detector. The micro-IR spectra of the ynamines powders were collected in transmission mode by depositing the samples on a diamond anvil cell (DAC). IR spectra of the solutions were collected at room temperature through a cell for liquid samples with KBr windows employing a Nicolet Nexus FTIR (4 cm$^{-1}$ resolution) equipped with a DTGS detector. 32 scans (4 cm$^{-1}$ resolution) were collected for each reported IR spectrum.

## 3. Calculation

Geometry optimization, the calculation of IR/Raman spectra and the first hyperpolarizability of the molecular models reported in Fig. 1 (isolated single molecules) have been carried in the framework of Density Functional Theory (DFT) by using GAUSSIAN09 package[47]. Calculations have been carried out at PBE0/cc-pVTZ level of theory, previously adopted for many other polyyne structures to obtain a nice agreement with experimental spectra[6,20]. The computed vibrational frequencies have been scaled by a factor of 0.9578, determined by adjusting the phenyl stretching mode predicted by DFT at ≈1667 cm$^{-1}$ to the peaks observed at ≈1597 cm$^{-1}$ in solution samples.

## 4. Results and discussion

### 4.1. The IR and Raman spectra of ynamines

We focus on the A[n] series, taken as representative of all the ynamines samples investigated in this work. The spectra of the other polyynes are reported in the Supporting Information (SI), i.e. the B[n] series in Fig. S1, C[n] in Fig. S2, and D[n] in Fig. S3.

The IR and Raman spectra are shown in Figure 2 for both solution and powder samples and chain length from 4 to 8 sp-carbon atoms ([n] as the number of acetylenic units moving from 2 to 4). Experimental results are compared with DFT simulations carried out on isolated molecules. The Raman spectra are normalized to the intensity of the phenyl ring stretching mode located at about 1600 cm$^{-1}$ and are dominated by the intense ECC peak in the 2000 to 2200 cm$^{-1}$ spectral region. The increased conjugation results in a remarkable increase of the intensity of the ECC band with sp chain length, shown in Fig. 2. Such increase is even larger than what is shown in Fig. 2 due to the adopted normalization. Indeed – as shown by theoretical predictions (see Table 1) – also the phenyl band, selected as an internal reference, is affected by the increase of the conjugation length and its Raman intensity growths with n.

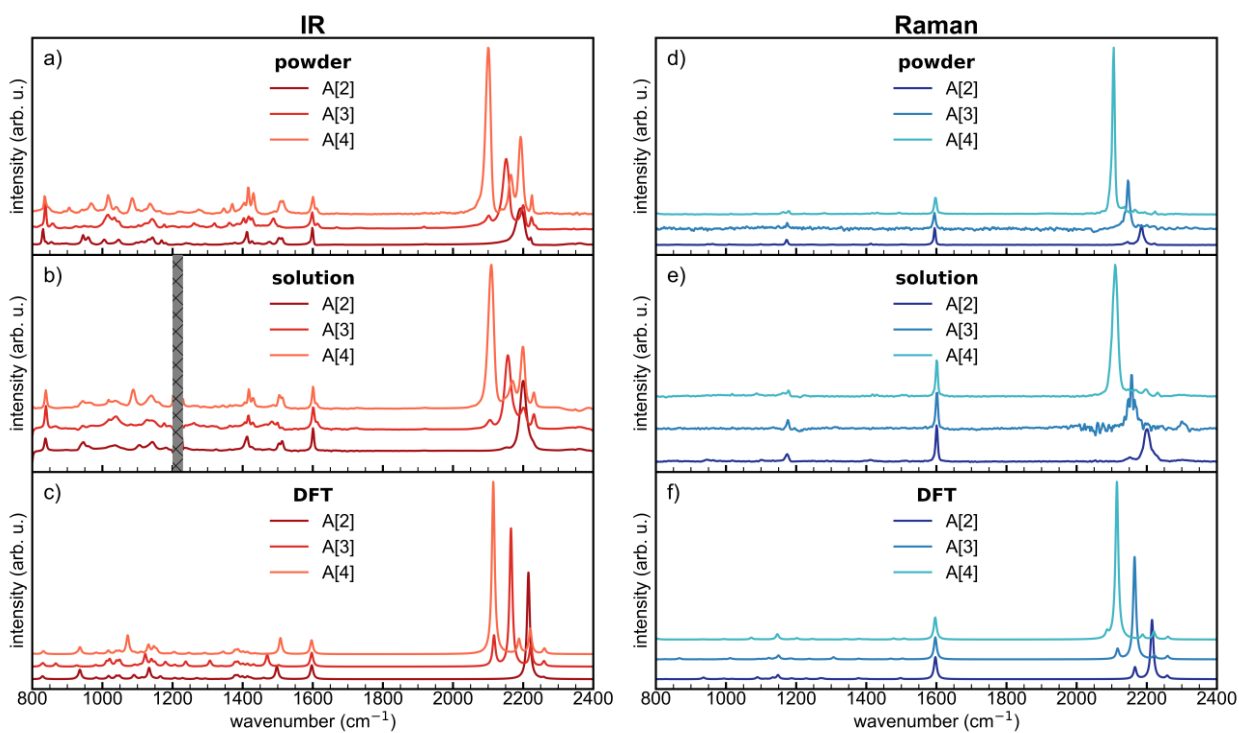

***Figure 2.*** *Panels a) and d) report powder IR and Raman spectra of the A[n] series (n=2-4), panels b) and e) display IR and Raman solution spectra and panels c) and f) show IR and Raman DFT calculations, respectively. The grayed region in panel b) is not measurable because of the strong IR absorption of the solvent.*

As confirmed by the analysis of the vibrational eigenvectors computed by DFT, the strongest Raman lines correspond to the ECC mode of the polyyne chain, i.e. the collective in-phase stretching and shrinking of the C≡C and C-C bonds[13,20]. This peculiar vibrational mode is characteristic of π-conjugated materials and coincides with the oscillation of the bond length alternation (BLA) parameter[20,21]. As expected, the ECC band shows a pronounced frequency dispersion and intensity modulation directly related to the length of the sp chain. Differently from the case of apolar or slightly polar polyynes, the ECC Raman transition is very intense also in the IR spectra of the A[n] series, as well as in the other ynamines here investigated. Measurements were performed both in solution and in powder to observe solid-state effects. If we consider the spectra in solution, we observe a perfect coincidence in the peak position of the strongest IR band with the ECC line identified in the Raman spectra (Fig. 3). The corresponding DFT calculations, that mimic isolated molecules, predict this behavior, as displayed in Fig. 2 and illustrated by the theoretical data reported in Table 1.

|  | Vibrational wavenumber (unscaled) [cm$^{-1}$] | Vibrational wavenumber (scaled) [cm$^{-1}$] | Intensity IR [km/mol] | Raman Activity [A$^4$/amu] |
|---|---|---|---|---|
| **A[2]** | 1668 | 1597 | 297 | 5563 |
|  | 2262 | 2166 | 1 | 3974 |
|  | **2313** | **2215** | **2284** | **20712** |
|  | 2358 | 2258 | 25 | 1069 |
| **A[3]** | 1667 | 1597 | 240 | 11122 |
|  | 2210 | 2117 | 520 | 6676 |
|  | **2261** | **2165** | **2385** | **70103** |
|  | 2320 | 2222 | 354 | 2391 |
|  | 2359 | 2259 | 87 | 2262 |
| **A[4]** | 1667 | 1597 | 278 | 18500 |
|  | 2178 | 2087 | 1 | 6228 |
|  | **2208** | **2115** | **3465** | **175496** |
|  | 2285 | 2188 | 288 | 4920 |
|  | 2318 | 2220 | 500 | 8838 |
|  | 2360 | 2261 | 108 | 3175 |

*Table 1.* Calculated frequencies and intensities for the vibrational modes of the A[n] series (n=2-4) in the region 1600 – 2300 cm$^{-1}$. The values relative to the ECC mode are highlighted with bold characters.

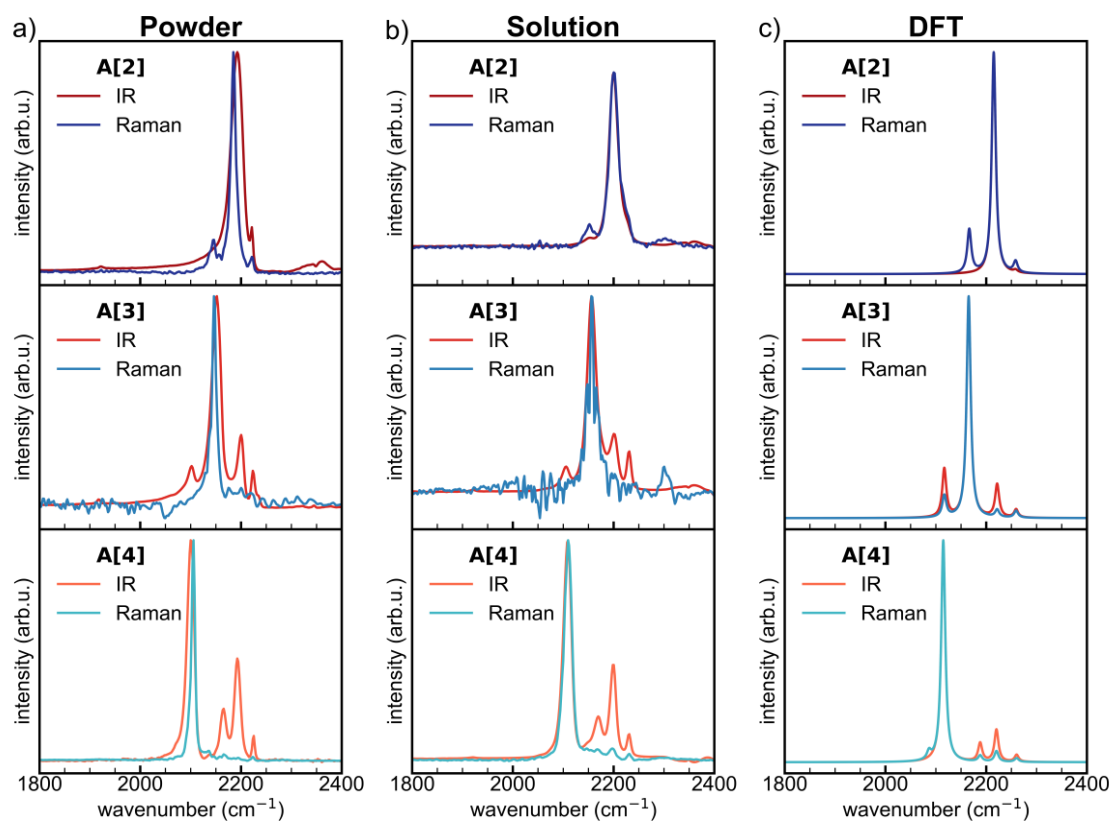

*Figure 3.* Comparisons of experimental IR (reddish tones) and Raman (bluish tones) spectra of the A[n] series (n=2-4) both in (a) powder and (b) solution, in the region dominated by the ECC mode.

Interestingly, a small but non-negligible shift (about 4 cm$^{-1}$) of the main ECC line is observed by comparing the experimental IR and Raman spectra of solid-state samples, as shown in Fig. 3. This is an indication that the Raman active modes are slightly affected by intermolecular interactions: molecules in the crystal may form pairs oriented with antiparallel dipoles with inversion center symmetry. Such a kind of molecular packing is observed in the solid-state of A[3] and D[4]: the crystal structure of A[3] is available in the work of Pigulski et al.[29], while that of D[4] is reported in Fig. S15 in the SI and it represents the first X-ray structure of nitrogen-capped octatetrayne. The vibrational modes of the single-molecule give rise to two different modes of the pair, where two nearest molecules vibrate in-phase (*gerade*, Raman active) or out-of-phase (*ungerade*, IR active). The intermolecular interactions determine the frequency splitting between such *gerade* and *ungerade* vibrations of the crystal, and the symmetry selection rules determine the mutual exclusion of the pair of vibrational modes in the IR and Raman spectra, as suggested by the non-coincidence of the strong ECC transition in the IR and Raman spectra of the powder samples (see Fig. 3).

In the 2000-2200 cm$^{-1}$ region, besides the previous intense signals, we can recognize other weaker vibrational transitions, both in IR and Raman spectra, whose number increases with the chain length. This feature was already encountered in other polyyne systems, where their phonon dispersion was investigated using an oligomer approach[13,22,68]. The stretching mode of the C≡N group of the benzonitrile termination shared by all ynamines is present in both Raman and IR spectra, located at ≈2220 cm$^{-1}$, regardless of chain length or termination. In the region between 800 and 1600 cm$^{-1}$, the IR spectra are characterized by crowds of signals that can be assigned to vibrational modes of the endgroups of ynamines.

The rationalization of the spectroscopic behavior of ynamines, especially in the 2000-2200 cm$^{-1}$ range, can be undertaken by considering the two main features of such molecules. From one side, with increasing chain length, ynamines behave like homo-terminated polyynes; from the other side, the observed activation in the IR of the ECC modes is the remarkable consequence of the push-pull nature of ynamines and deserves a careful discussion in connection with their large first hyperpolarizability.

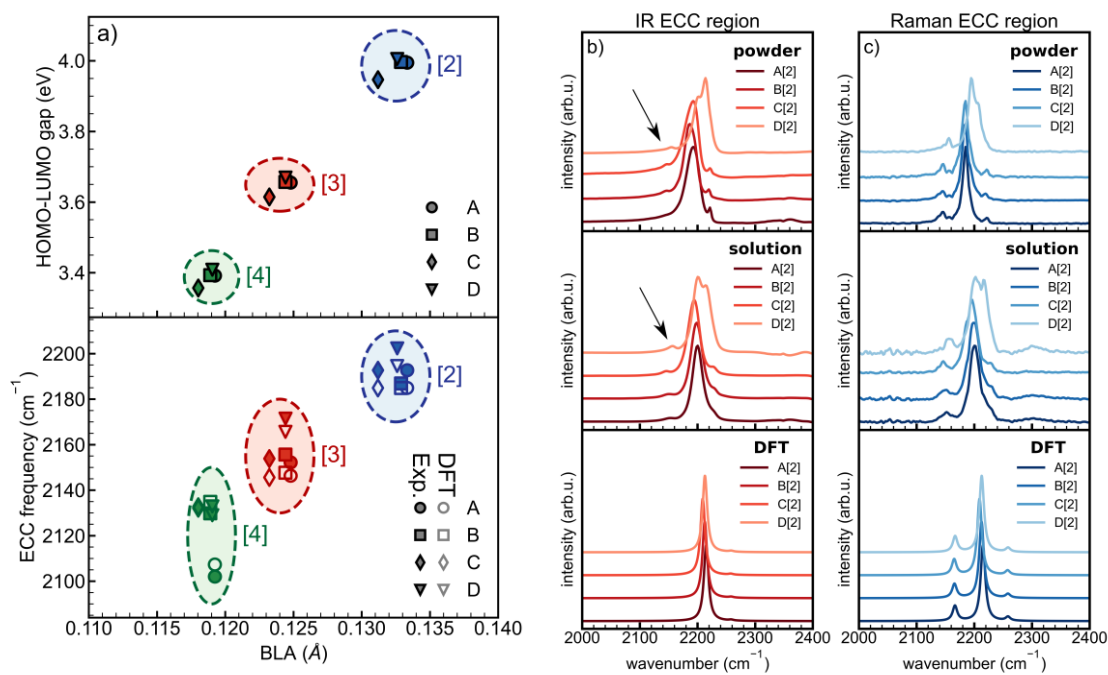

*Figure 4.* a) Evolution of the HOMO-LUMO gap and the ECC peak position as a function of BLA of ynamines. Each termination is labeled by a specific marker (circle for A, square for B, thin diamond for C, and triangle for D). Ynamines with the same number [n] of triple bonds are inserted into colored circles and marked with the corresponding tag (blue for [2], red for [3], and green for [4]). Experimental and DFT data are classified by filled and empty markers, respectively. b) panel reports the region between 2000 and 2400 cm$^{-1}$ of powder, solution and DFT calculated IR spectra, respectively, of ynamines with [2] triple bonds. c) panel reports the region between 2000 and 2400 cm$^{-1}$ of powder, solution, and DFT calculated Raman spectra, respectively, of ynamines with [2] triple bonds.

The properties of ynamines can be investigated following the approach generally adopted to analyze polyynes and other π-conjugated compounds. Figure 4 shows the correlations between the BLA parameter, the HOMO-LUMO gap calculated by DFT, and the mean position of the ECC peak obtained from the experimental IR and Raman spectra of powder samples and their counterparts predicted by DFT. Ynamines possess optical gap values in the near UV region (3.3 – 4 eV), that decrease with the increase of π-conjugation[5,16]. Similarly, the HOMO-LUMO gap decreases with BLA, since the increase of π-conjugation leads to a decrease of BLA. For the systems investigated here, it turns out that the BLA is mainly determined by the length of the polyyne bridge and less affected by the endgroups: ynamines with the same length and different endgroups exhibit quite similar BLAs and can be enclosed in a small region of the BLA *vs*. HOMO-LUMO gap plot. Similar considerations apply to the position of the ECC peaks as a function of BLA. We observe the expected bathochromic shift of the ECC position with decreasing BLA, or equivalently with the increase of the sp chain length [14,16,22,23,60,69,70]. Ynamines with the same size, display rather close positions of the ECC peaks, independent from the differences in the donor group. This effect is usually observed in size-selected polyynes with chemically similar and symmetric endgroups[14,16,60,69]. The substitution with a chemically different termination may introduce a measurable shift of the ECC position, as in the case of CN-capped polyynes, i.e. H-$C_n$-H *vs* H-$C_n$-CN[71,72]. Therefore, we have a spectroscopic indication that the different donor groups of ynamines similarly affect the π-conjugation. This is confirmed by the examination of panels b) and c) of Fig. 4, where are presented zoomed regions (2000-2400 $cm^{-1}$) of the experimental and simulated IR and Raman spectra of the ynamines belonging to the [2] series.

While the main spectral features of ynamines are well reproduced by DFT simulations, there are some discrepancies between theory and experiments. First, the infrared intensity of a mode located at about 2145 $cm^{-1}$ in all [2] series (see black arrows in Fig. 4b)) is underestimated by the calculations. This mode is associated with a collective vibrational mode of the sp chain involving BLA oscillation and showing a node on the central C–C bond (see Fig. S4 in the SI), as found in other polyynes[13,22–24,70]. Second, DFT calculations do not predict the shape of the ECC band of D[2]. This structure could be explained by a peculiar coupling between the C≡N stretching and ECC vibrations. Indeed D[2] is the molecule showing a higher ECC frequency, very close to that of the C≡N stretching mode. These discrepancies could be associated with the over-delocalization of the π-electrons described by DFT calculations, which predict C≡N and ECC modes too far apart and overestimate the collective character of the 2145 $cm^{-1}$ mode.

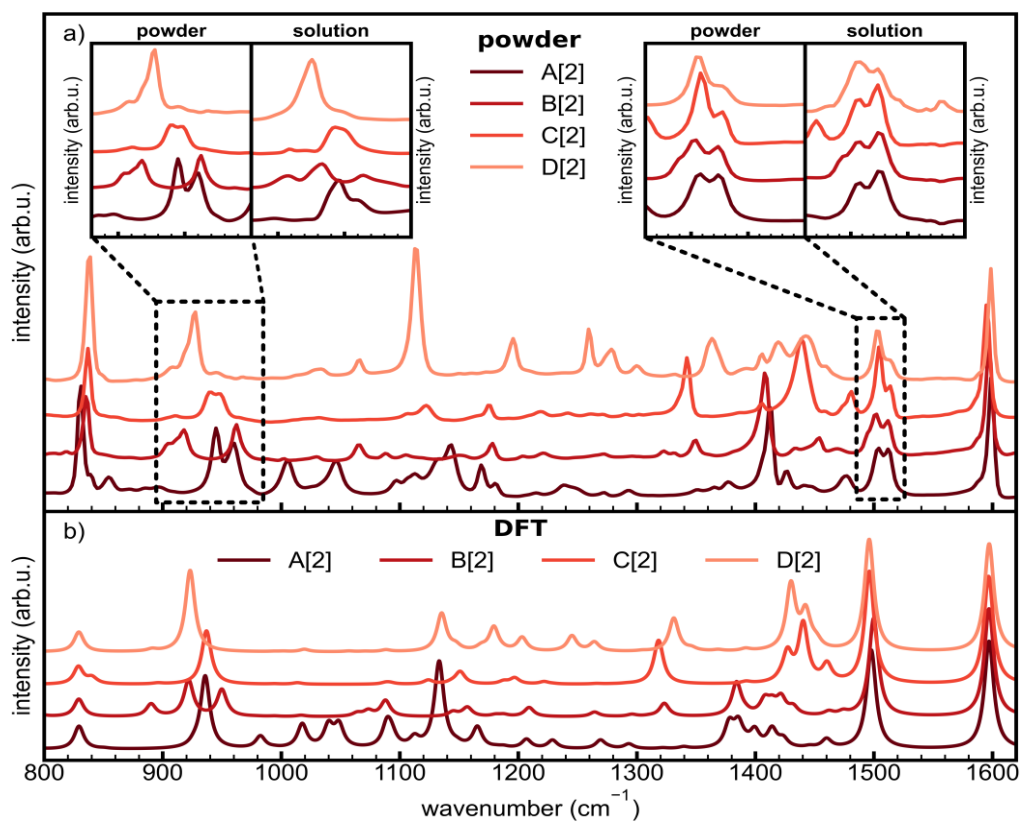

*Figure 5.* Focus on the spectral region between 800 and 1600 cm$^{-1}$ of the IR spectra of the series [2] of ynamines. a) Experimental IR spectra of powder samples. The two insets report slices of the IR spectra of solutions, in the left panel between 880 and 1100 cm$^{-1}$ and the right inset between 1480 and 1550 cm$^{-1}$. b) DFT calculations of the same ynamines of panel a).

The region of the IR spectra of ynamines that ranges from 800 to 1600 cm$^{-1}$, reported in Figure 5, is rich of bands. In this region, the IR spectra of ynamines with the same length slightly differ by changing the donor group. Therefore, this region of the IR spectrum can be used to identify some markers of the specific donor endgroups. In SI a detailed discussion of these features is illustrated (see Section "Donor-dependent markers in the mid-low IR spectrum" in the SI).

The analysis of the experimental spectrum in this region put into light several weaknesses of DFT calculations carried out on individual isolated molecules: experiments show the presence of two rather strong doublets centered around 950 cm$^{-1}$ and around 1500 cm$^{-1}$ respectively (see insets in Figure 5), where the predicted spectrum shows one only strong IR transition; moreover, at lower wavenumbers, the predicted spectrum does not fit well the experimental spectra both of powders and solution, both showing a very strong transition at about 830 cm$^{-1}$, not reproduced by the theory. The appearance of doublets in the experimental spectra of crystalline phases is often ascribed to crystal field splitting, that, however, is not expected in our case because of the presence of an inversion center, which relates the two molecules forming the basis of the crystal (P$\bar{1}$ space group). Interestingly, also molecules in solutions show similar features, even if the components forming the doublets have different relative intensities compared to the crystalline phases.

Periodic boundary conditions DFT simulation performed on the crystal of the A[3] molecule shows a better agreement with the experimental data. Compared to single-molecule calculation, it shows a larger number of modes with remarkable IR intensity in the mid-low region of the IR spectra (see Fig. S7 in the SI). The computed vibrational eigenvectors show that this behavior can be ascribed to intermolecular interactions that scarcely modify the vibrational dynamics (and frequencies) but determine mutual polarization effects on the interacting molecules. This results in a remarkable activation of some IR modes, which were practically silent for the isolated molecule. Also, the strong feature at about 830 cm$^{-1}$ is well reproduced by the calculations for the crystal, showing in this region several intense IR transitions associated with complex collective vibrational modes. In this case, the strongly IR active modes are characterized by a mixing of ring breathing and out-of-plane CH bending modes, that do not find a counterpart among the normal modes of the isolated molecule. While the DFT calculation of the crystal allows us to explain observed features which are not captured by single-molecule calculations, we still need to explain why the experimental spectra of ynamines in solutions (in the insets of Fig. 5a) show similar features as those observed in the solid-state. A way to rationalize these findings is considering that a non-negligible population of clusters and/or dimers survived in all the solutions, together with single molecules. In fact, the theoretical prediction of the vibrational spectra of a dimer of the A[3] molecule (starting geometry "extracted" considering closely packed dimers occurring in the crystal) nicely supports the above conclusion (see Fig. S7 in the SI). The vibrational spectrum predicted for the dimer shows features correlated to the pattern obtained from simulation of the crystal; overall, the IR spectrum of the dimer is more structured than that of the single-molecule and seems to anticipate the relevant effect observed for the crystal, namely the activation in the IR of some modes with vanishing intensity in the spectrum of the isolated molecule.

To experimentally check the soundness of our interpretation, spectra of the solution at variable concentration spectra of molecule A[2] have been recorded (see Figure S6 in the SI). All the spectra show doublets around 950 cm$^{-1}$ and 1500 cm$^{-1}$, even at the lowest concentration, i.e. 8 x 10$^{-4}$ M, near the detection limit of our instrument. Remarkably, we can observe a slight evolution of the band shapes (relative intensities of the components of the doublets) toward the powder sample spectrum, while increasing the concentration. Even if these finding, together with simulations of the dimer, suggests that intermolecular interactions play a significant role also in diluted solution, there is another phenomenon that justifies some weakness in the prediction of the solution spectra by DFT simulations.

As illustrated in SI (Fig. S8) for case A[4], the presence of "flexible" bonds belonging to the donor group, gives rise to several stable conformations of the molecules. The contribution to the solution spectrum from few low-energy conformers could justify the presence of additional bands which are not present in the single-molecule simulations.

## 4.2. IR activation of the ECC mode as a marker of π-electrons polarization

The coincidence of Raman and IR bands in polyynes has been already reported in the case of slight deviations from linearity induced by the bending of the sp chain or in the case of uneven terminations, that cause the breaking of the local inversion symmetry[22–24,73]. In those cases, the symmetry-allowed vibrational modes of IR spectra were weakly activated in the Raman spectra, and vice versa, due to the dynamic polarization of the sp carbon chain[22–24]. DFT simulations could rationalize the observed violation of the IR/Raman exclusion principle for symmetrically substituted polyynes in presence of slightly bent polyyne models, while calculations cannot predict this phenomenon in the case of asymmetric terminations. The case of ynamines is different and it can be analyzed in the framework of push-pull π-conjugated molecules. The structure of ynamines is similar to that of push-pull polyenes, in which the π-conjugated skeleton is made by $sp^2$-hybridized carbon atoms[45,46]. Push-pull polyenes also feature a strong IR signal assigned to the ECC mode of the π-conjugated bridge, the origin of which has been widely investigated in the past[45,46].

The evolution of the infrared intensity of the ECC band in the IR spectra of the A[n] series of ynamines as a function of the number of triple bonds is reported in Figure 6. The experimental intensity values, calculated as the integrated area of the band and normalized to the total CH stretching intensity (3150-2750 $cm^{-1}$), show a monotonic growth of the IR intensity of the ECC band, with the length of the sp carbon chain. The selected CH stretching region is a good internal reference, because it shows a very stable infrared intensity along with the A[n] series, independently on the polyyne chain length. Figure 6 shows a similar trend of experimental and calculated intensities. The calculated IR intensities slightly increase from A[2] to A[3], while a relevant increase occurs for A[4], differently from the dipole moment values showing a linear increase with chain length. This jump from A[3] to A[4] appears to be slightly overestimated by DFT calculation.

The explanation of the observed IR intensity trend can be based on a simple model which uses the concept of "effective internal electric field F", induced by the polar terminations and affecting electrons belonging to the chain[46]. Based on this model, a linear relationship between the Raman and IR intensities of ECC modes of polyenes is expected[45], whose proportionality constant is the square of the effective internal field, $F^2$. Moreover, under certain conditions, the effective field in the dipole direction can be easily estimated as the ratio of the static dipole moment and the molecular polarizability, i.e. $F = \frac{\mu}{\alpha}$, where $\alpha$ is the diagonal component of the static molecular polarizability in the dipole direction. This model can explain the behavior of the IR intensities of ynamines. Despite the linear increase of the molecular dipole moment, F decreases from A[2] to A[4], because of the leading role of the denominator, namely of the static molecular polarizability that increases more than linearly with the sp carbon chain length. Conversely, the Raman intensity of the ECC mode increases markedly with chain length. The opposite trends of F and the ECC Raman intensity determine the observed behavior of IR intensity with n.

On the other hand, the linear increase of the dipole moment (Fig. 6) is explained if the charge transfer between (D) and (A) groups is supposed to be independent of the polyyne length, simply considering the effect of the increasing distance between (D) and (A) groups, while increasing the $\pi$ bridge length. Interestingly, this observation tells us that the whole chain is involved in the charge transfer process, thus indicating that we are far from the typical "saturation" of the push-pull behavior observed for long polyene systems[38,74].

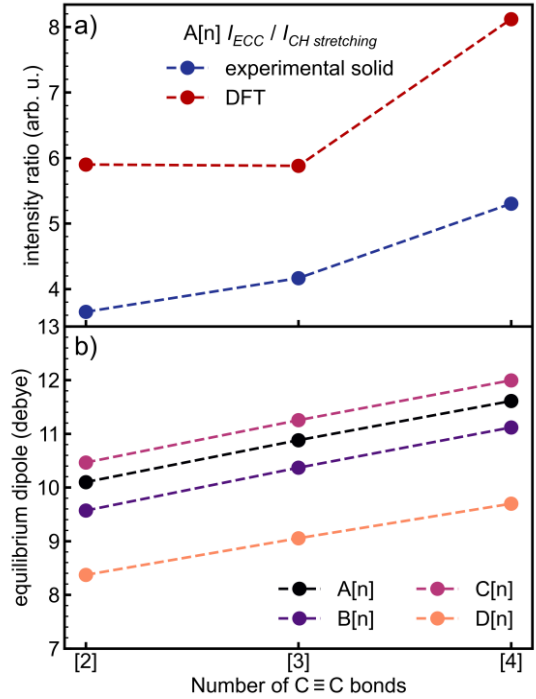

*Figure 6.* a) Evolution of the integrated intensity of the ECC IR band normalized to the CH stretching total intensity from the experimental spectra of Fig. 2a) and from DFT calculations b) predicted dipole moment for the series A[n] of ynamines.

### 4.3. First hyperpolarizability: vibrational and electronic contributions.

The vibrational first hyperpolarizability ($\beta^v$) is usually large in push-pull molecules and it can be expressed as follows:

$$\beta^v_{nmp} = \frac{1}{4\pi^2 c^2} \sum_k \left(\frac{1}{v_k^2}\right) \times \left[\frac{\partial \mu_n}{\partial Q_k}\frac{\partial \alpha_{mp}}{\partial Q_k} + \frac{\partial \mu_m}{\partial Q_k}\frac{\partial \alpha_{np}}{\partial Q_k} + \frac{\partial \mu_p}{\partial Q_k}\frac{\partial \alpha_{mm}}{\partial Q_k}\right] \quad \text{(Eq. 1)}$$

where the index k denotes a given vibrational transition, with associated wavenumber $\nu_k$, normal coordinate $Q_k$, dipole derivative $\frac{\partial \boldsymbol{\mu}}{\partial Q_k}$ and polarizability derivative $\frac{\partial \boldsymbol{\alpha}}{\partial Q_k}$. Both derivatives in Eq. 1 are evaluated at the equilibrium geometry. Very often, to ease comparisons, the vector component of the $\beta$ tensor is introduced:

$$\beta_\mu = \frac{\sum_{i=x,y,z} \mu_i(\beta_{ixx} + \beta_{iyy} + \beta_{izz})}{\sqrt{\mu_x^2 + \mu_y^2 + \mu_z^2}} \quad \text{(Eq. 2)}$$

Since IR intensities and Raman activities depend on dipole derivatives and polarizability derivatives, respectively, vibrational spectra can provide the quantities required for the evaluation of $\beta^v$. The procedure for the experimental determination of $\beta^v$ is illustrated in Castiglioni et al.[56] and it has been adopted for the characterization of the NLO response of several organic molecules. The determination of $\beta^v$ can be easily obtained by Quantum Chemical calculations using the same Eq. (1) which shows that large $\beta^v$ values can be obtained when for the same normal coordinate $Q_k$, both $\frac{\partial \boldsymbol{\mu}}{\partial Q_k}$ and $\frac{\partial \boldsymbol{\alpha}}{\partial Q_k}$ are large. This may happen when the molecule displays vibrational transitions that are strong both in IR and Raman. Therefore, the large intensity of the ECC transition observed both in the Raman and in the IR spectra of ynamines suggests potentially large $\beta^v$ values. Table 2 confirms the sizeable values (computed by DFT methods) of the vibrational and electronic first hyperpolarizabilities of ynamines.

Our data reported in Table 2 and Figure 7, indicate that the order of magnitude of $\beta$ is $10^{-28}$ esu, within the typical range of organic molecules designed for NLO applications. We can notice that the vibrational hyperpolarizability ($\beta^v$), calculated for all the ynamines here considered, provides a good estimate of the electronic counterpart ($\beta^e$) since the $\beta^v$ values are between 80% and 90% of $\beta^e$. Moreover, the electronic and vibrational hyperpolarizabilities increase upon increasing the number of triple bonds in the sp carbon chain, as observed in many other π-conjugated systems[57,62,75]. The strong IR and Raman active ECC mode plays a remarkable role in determining the large $\beta^v$ values of ynamines. As reported in Table 2, the ECC mode is always responsible for more than 50% of the total $\beta^v$ value. Moreover, the relative importance of the contribution of the ECC mode to $\beta^v$ increases with the chain length, reaching about 70% for n = 4 triple bonds. This finding proves that (D) and (A) groups of ynamines interact through the π bridge at all chain lengths and the polarization induced by the donor-to-acceptor charge transfer involves the whole polyyne chain. This interpretation is supported by the fact that the ECC transition is largely dominant both in the IR and in the Raman spectra, and the ECC mode involves nuclear displacements over all the CC bonds of the sp carbon chain. The leading role of the vibrational intensities of the ECC mode in $\beta^v$ is even more remarkable considering that in Eq. (1) the large ECC vibrational frequency shows up at the denominator and has therefore a damping effect on the ECC contributions.

The reliability of the hyperpolarizability values obtained by DFT calculations, which are known to overestimate the effects of π electron delocalization, should be carefully considered[13]. The limits of DFT calculations in the description of some minor spectroscopic features may suggest that the physical quantities related to the π electrons response could be overestimated. To get more indications, we computed the hyperpolarizabilities of ynamines by the Hartree-Fock (HF) method (see Table S1 and S2 in SI), which is known to provide a more localized description of π electrons than DFT[13]. The vibrational and electronic hyperpolarizabilities computed by HF are lower than those computed by DFT, but they are of the same order of magnitude. Moreover, the increase of hyperpolarizability with chain length is less steep in HF calculations than in DFT. However, the spectra predicted by HF are poor in the description of the observed spectral pattern of ynamines: HF calculations cannot reproduce the distinctive spectroscopic signature of such push-pull polyynes, where the ECC transition dominates both Raman and IR spectra. This observation, corroborated by the conclusions reached in de Wergifosse and Champagne[63] about the reliability of DFT for the calculation of the hyperpolarizability of push-pull polyynes up to 4 triple bonds, convinced us that we can be rather confident on the DFT results for these short polyynes. This conclusion is further supported by the calculations carried out for the series A[n], reported in SI. Vibrational and electronic first hyperpolarizabilities have been calculated also adopting B3LYP DFT functional and 6.31G** basis set, following the indication of de Wergifosse and Champagne[63]: the values obtained show a good agreement with the same parameters calculated according to the PBE0/cc-pVTZ level of theory, adopted in our whole investigation. However, the results obtained here with DFT methods for extrapolating the β values of longer chains are not recommended[69].

|  | A | | | B | | | C | | | D | | |
|---|---|---|---|---|---|---|---|---|---|---|---|---|
|  | [2] | [3] | [4] | [2] | [3] | [4] | [2] | [3] | [4] | [2] | [3] | [4] |
| $\beta^v$ | 0.632 | 1.113 | 1.744 | 0.698 | 1.209 | 1.863 | 0.614 | 1.100 | 1.741 | 0.574 | 1.022 | 1.603 |
| $\beta^e$ | 0.760 | 1.331 | 2.055 | 0.772 | 1.362 | 2.110 | 0.769 | 1.350 | 2.089 | 0.754 | 1.296 | 1.976 |
| $\beta^v_{ECC}$ | 0.289 | 0.567 | 1.134 | 0.316 | 0.611 | 1.231 | 0.307 | 0.594 | 1.206 | 0.294 | 0.564 | 1.113 |

**Table 2** Vibrational and electronic $\beta_\mu$ values (units of $10^{-28}$ esu) for A[n], B[n], C[n] push-pull polyynes, from DFT calculations. $\beta_\mu$ values, obtained considering the only contribution of the ECC band, are also reported. In the calculation of $\beta^v$ we considered only modes with wavenumber above 100 cm-1, according to refs. [39] and [45][56,57].

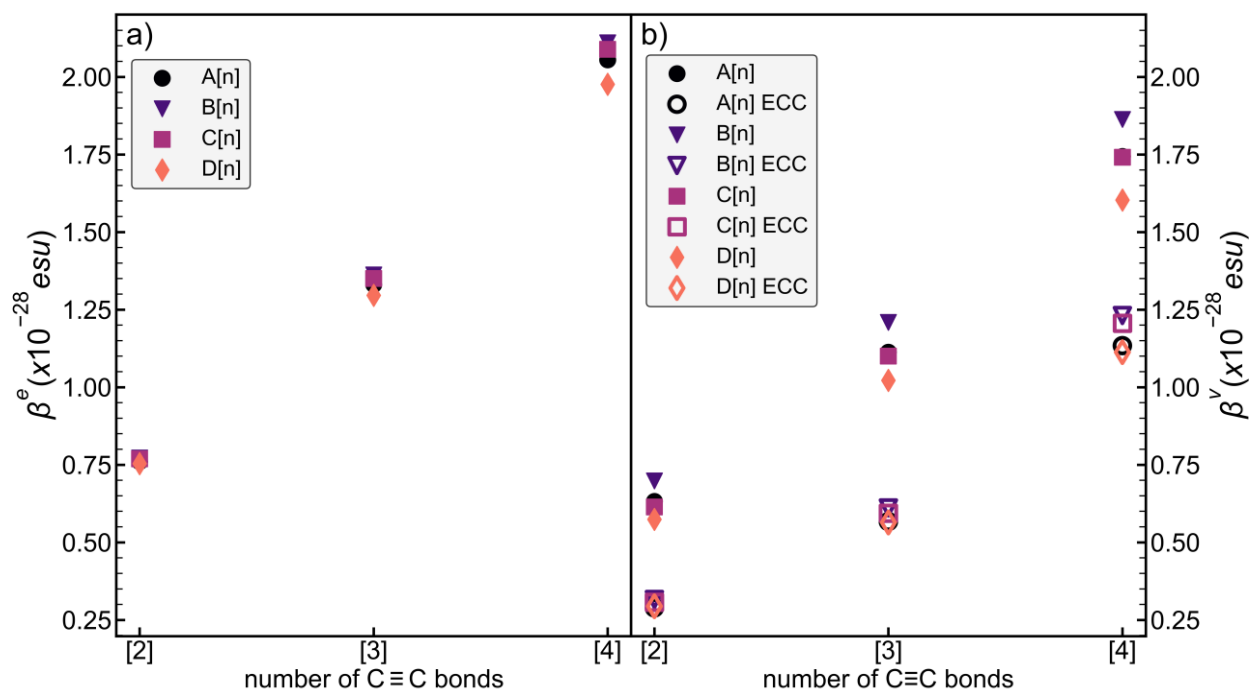

*Figure 7.* Left panel: Electronic $\beta_\mu$ values (units of $10^{-28}$ esu) for A[n], B[n], C[n] push-pull polyynes from DFT calculations. Right panel: vibrational $\beta_\mu$ values (units of $10^{-28}$ esu) for A[n], B[n], C[n] push-pull polyynes from DFT calculations (solid symbols) and $\beta_\mu$ values obtained considering the only contribution of the ECC band (open symbols).

## 5. Conclusions

We have performed a detailed analysis of the vibrational properties of ynamines that represent one of the examples of stable push-pull polyynes. Their polar donor and acceptor endgroups connected through polyyne bridges generate a strong dipole moment along the direction of the sp carbon chain, like other push-pull systems. The intense IR peaks recorded in all ynamines for the collective vibrational mode of the polyyne skeleton (i.e., the ECC mode), prove that the π-conjugated carbon chain supports an effective charge transfer between the donor (D) and acceptor (A) endgroups. The experimental finding parallels DFT calculations that predict both the strong dipole moment and the enhancement of the IR signals. Ynamines also show the usual vibrational properties of polyynes, such as the shift of the ECC peak, BLA, and HOMO-LUMO gap as sp chain length increases. As for the push-pull behavior, in the four series of ynamines here investigated, characterized by a common acceptor endgroup, the experimental data, and DFT calculations confirm the similarity of the various donor groups. The IR spectra in the region between 800 and 1600 cm$^{-1}$ display interesting spectral markers of the different donor groups. Finally, the intense Raman and IR activities suggest appealing nonlinear optical properties that may be exploited in the field of optical communications and signal processing. Electronic and vibrational hyperpolarizabilities, computed through DFT simulations, support these conclusions, showing β values of the order of magnitude of $10^{-28}$ esu, with an increasing trend with chain length. The data here presented indicate that ynamines are attractive push-pull candidates for NLO applications. To this aim, some issues should be considered. As illustrated in Figure 3, the comparison between the IR and Raman spectra of ynamines in the solid-state shows a mismatch of the position of the ECC transitions in the two spectra: this is expected due to the formation of dimers related by inversion symmetry which leads to the mutual exclusion between IR and Raman transitions and implies that each term in Eq. 1 is zero. Hence the crystal has a vanishing first hyperpolarizability – both vibrational and electronic – due to the cancellation of the large β values of the individual molecules which form pairs in the centrosymmetric crystals. This occurrence is very frequent in large dipole molecules, which often pack in the solid-state with antiparallel dipoles. Like many other push-pull systems[38], to profit from the large molecular β of push-pull ynamines suitable strategies are needed to prevent centrosymmetric packing. This goal may be reached, *e.g.* by dispersing in polymer matrices the active molecules that can be easily poled by strong electric fields, or by suitable functionalization that makes energetically unfavorable the packing motifs possessing inversion symmetry.

## Acknowledgments

P.M., A.M., A. L., M. T., C. C. and C.S.C. acknowledge funding from the European Research Council (ERC) under the European Union's Horizon 2020 research and innovation program ERC-Consolidator Grant (ERC CoG2016 EspLORE grant agreement no. 724610, website: www.esplore.polimi.it).

**Supporting Information:** IR and Raman spectra of the series B[n], C[n] and D[n], normal modes representations, variable concentrations IR spectra of the molecule A[2], conformer IR spectra of the molecule A[4], DFT calculations of IR and Raman spectra of A[3], vector components of the first hyperpolarizabilities, synthesis and characterization ($^1$H and $^{13}$C NMR spectra) of the D[n] series, single crystal X-ray diffraction data of D[4].

# Supporting Information

*Pietro Marabotti†, Alberto Milani†, Andrea Lucotti‡, Luigi Brambilla‡, Matteo Tommasini‡, Chiara Castiglioni‡, Patrycja Męcik§, Bartłomiej Pigulski§, Sławomir Szafert§\*, C.S. Casari††*

† Micro and Nanostructured Materials Laboratory - NanoLab, Department of Energy, Politecnico di Milano via Ponzio 34/3, I-20133, Milano, Italy
‡ Department of Chemistry, Materials and Chem. Eng. 'G. Natta', Politecnico di Milano Piazza Leonardo da Vinci 32, I-20133, Milano, Italy
§ Faculty of Chemistry, University of Wrocław, 14 F. Joliot-Curie, 50-383 Wrocław, Poland

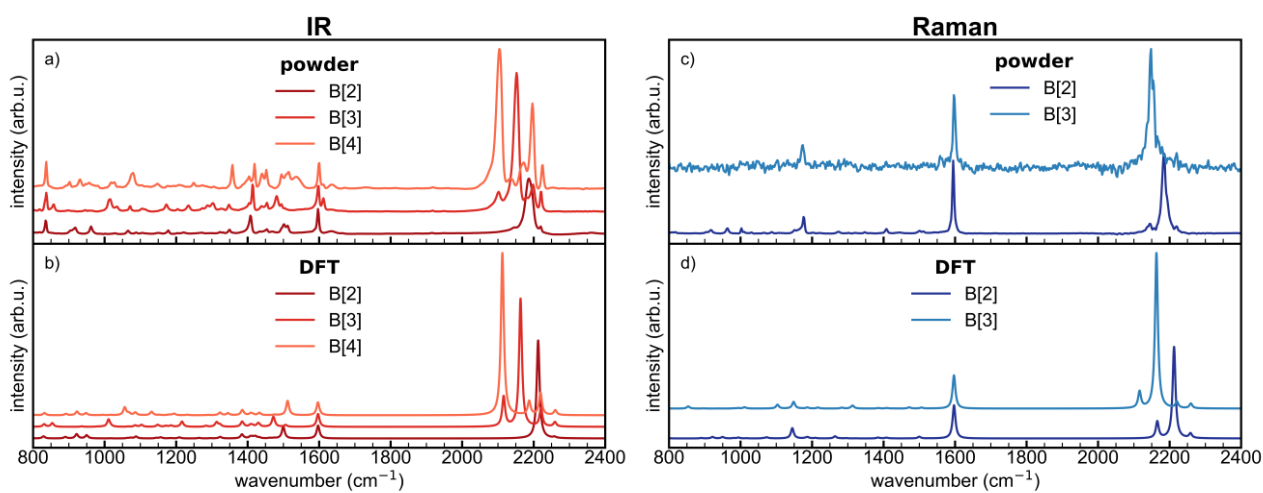

***Figure S1*** *Panels a) and c) report powder IR and Raman spectra of the B[n] series (n=2-4) and panels b) and d) show IR and Raman DFT calculations, respectively. The experimental Raman spectrum of B[4] cannot be recorded due to sample degradation.*

---

† *Corresponding authors: Tel. +39 02 2399 6331* Email: carlo.casari@polimi.it
Tel. +48 71 375 71 22 Email: slawomir.szafert@chem.uni.wroc.pl

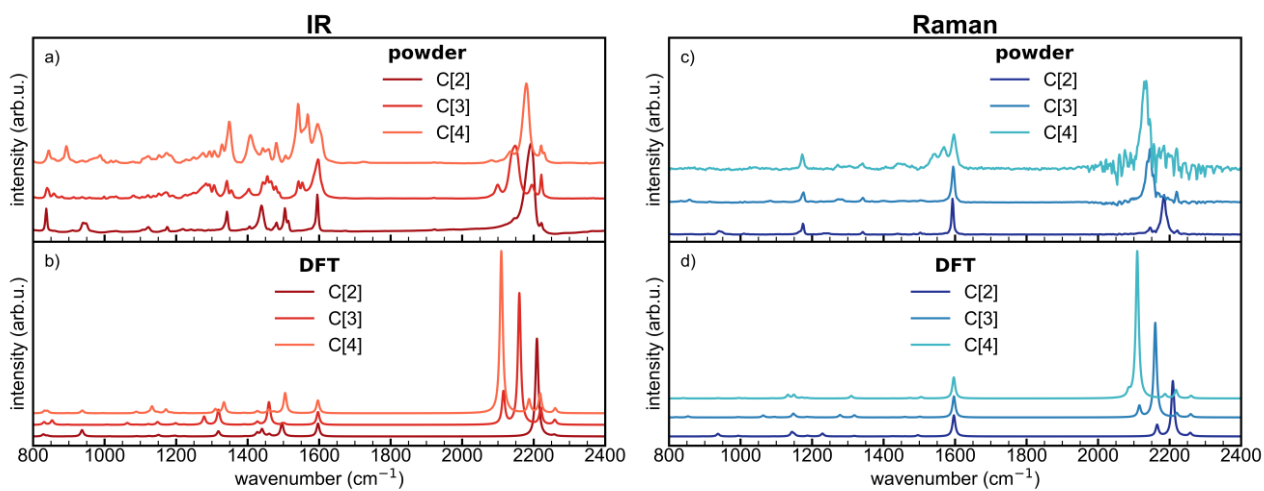

***Figure S2*** *Panels a) and c) report powder IR and Raman spectra of the C[n] series (n=2-4) and panels b) and d) show IR and Raman DFT calculations, respectively.*

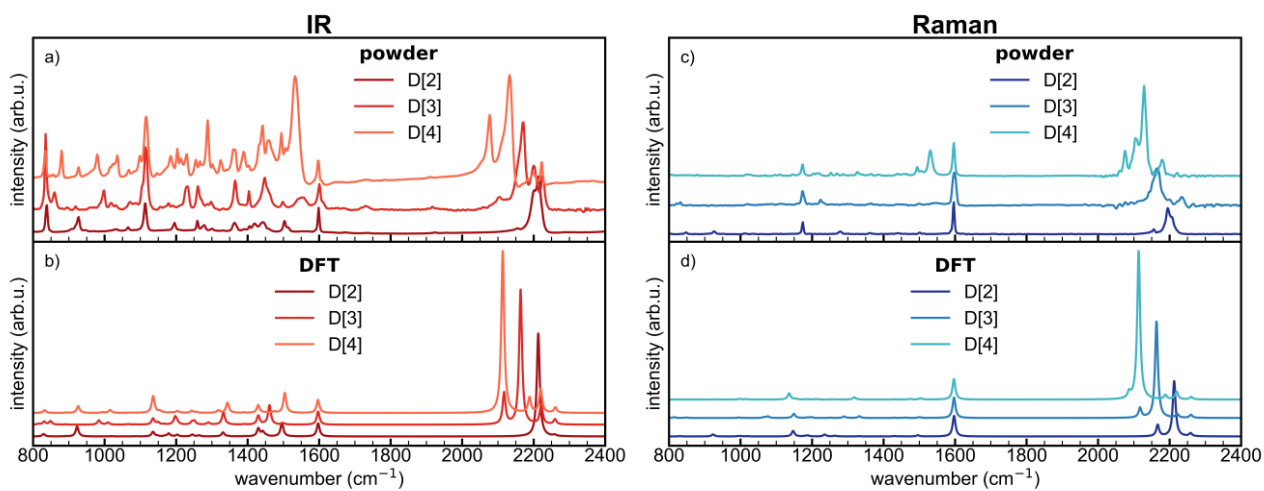

***Figure S3*** *Panels a) and c) report powder IR and Raman spectra of the D[n] series (n=2-4) and panels b) and d) show IR and Raman DFT calculations, respectively.*

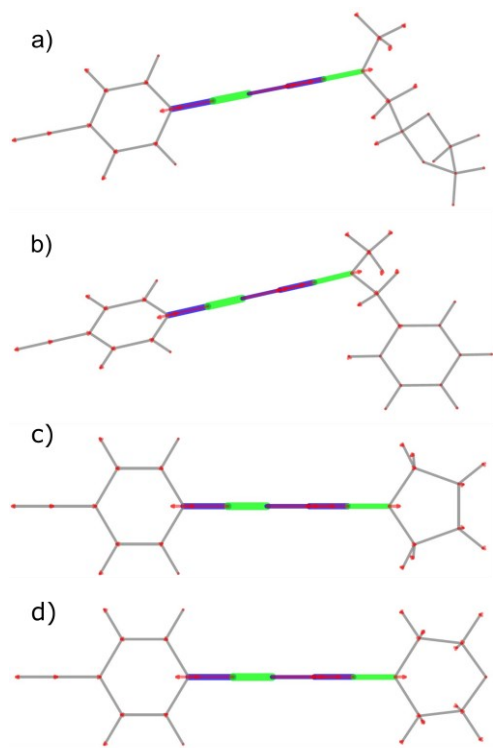

***Figure S4*** *Normal modes of the infrared signal centered around 2145 cm$^{-1}$ for all the [2] ynamines (from the top to the bottom: A[2], B[2], C[2] and D[2]).*

## Donor-dependent markers in the mid-low IR spectrum

Figure 5 (main text) shows spectra and DFT simulations of the powder samples of the [2] series. Starting from the low-frequency region, below 1000 cm$^{-1}$, only B[2] shows a remarkable difference in the vibrational modes characterized by out-of-plane bending of C-H bonds mainly localized on the donor group (see Fig. S5a). A[2], instead, possesses several active IR modes between 950 and 1150 cm$^{-1}$, where the other ynamines are quite silent, as predicted also by calculations (see Fig. 5b in the main text). These modes are localized on the A[2] donor group and they are characterized by CO stretching and CH bending (see Fig. S5b for the representation of the nuclear displacements of the two most intense peaks). In the region between 900 and 950 cm$^{-1}$, intense bands with doubled peaks can be mainly assigned to the C-N stretching of the donor termination, coupled with the stretching of the central C-C bond of the sp carbon chain (see Fig. S5c). Similarly, at about 1500 cm$^{-1}$, the experimental spectra show a pair of peaks in correspondence with the only one strongly IR active mode predicted by DFT calculations. This mode is assigned to the stretching of the central C-C bond and the shrinking of the C-N and C-C bonds at the edges of the polyyne bridge, coupled with a benzonitrile ring deformation (see Fig. S5d).

Considering the spectral region that extends from 1300 to 1450 cm$^{-1}$ in Fig. 5 (main text), we find – below 1350 cm$^{-1}$– two nearby peaks for C[2] and D[2]. They are characterized by significant C-H bending of the donor pentagon or hexagon ring, respectively (see Fig. S5e). Indeed, the endgroups of C[2] and D[2] are structurally quite similar, as can be seen in Fig. 1 (main text), which explains their comparable intensities and peak positions. From 1350 to 1450 cm$^{-1}$, we can classify the spectra into A-B and C-D couples. Indeed, the modes affecting this region are delocalized throughout the whole molecule, but the analogies of the donor terminations of A-B and C-D ynamines, respectively, resulting in a similar infrared pattern. All these normal modes have in common the stretching of the central C–C bond and the stretching of the C–C and C-N bonds at the edges of the sp carbon chain. The A-B pair presents modes with important CH bending on the donor termination, while the C-D group features donor and acceptor rings deformations, as pictured in Fig. S5f. Therefore, these bands are markers of the specific donor group structure of ynamines.

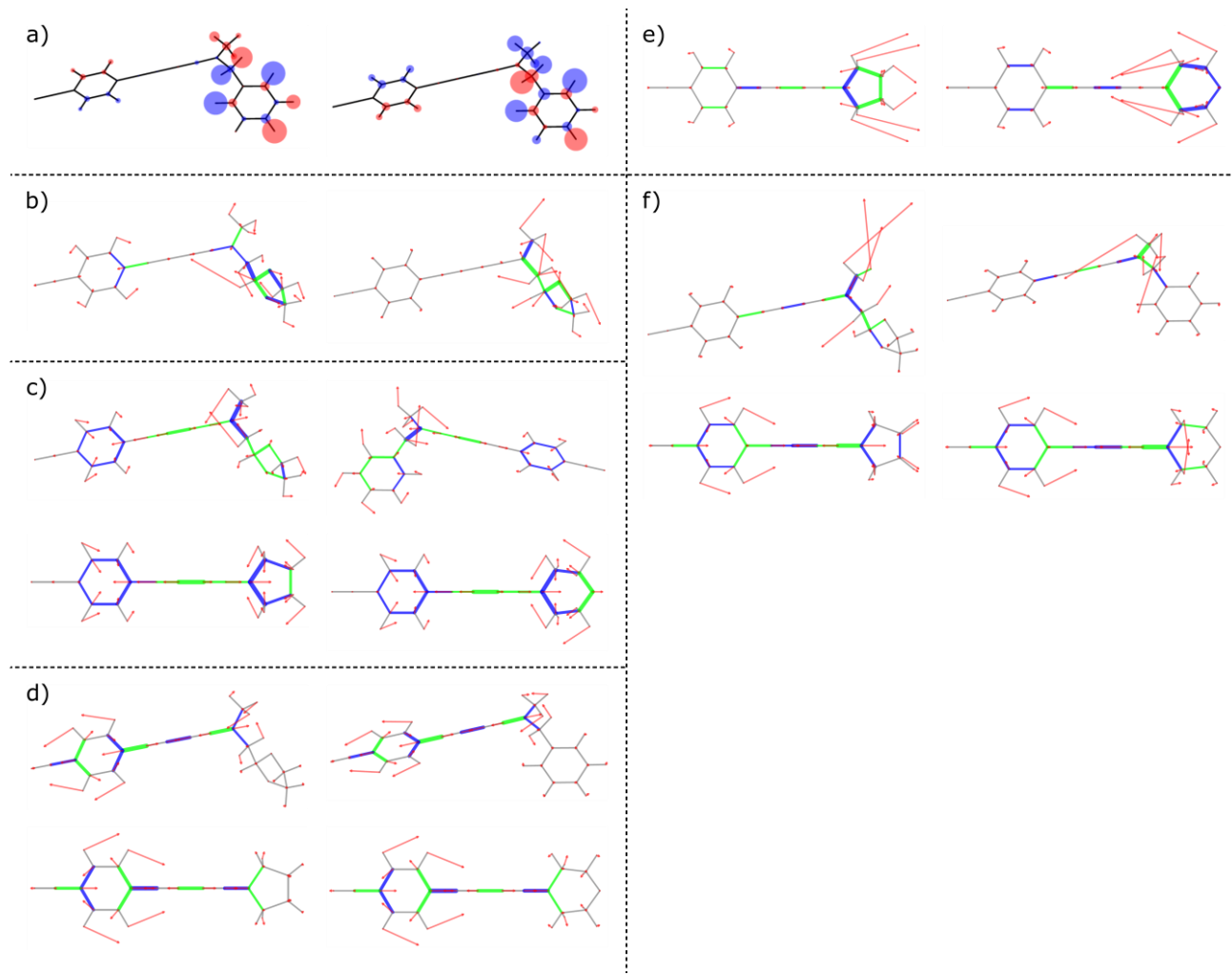

*Figure S5* Eigenvectors of marker bands in the mid-low region of the IR spectra of [2] ynamines. a) Out-of-plane bending of C-H bonds of the B[2] molecule at 890 and 922 $cm^{-1}$. b) C-O stretching and C-H stretching of the A[2] molecule, localized at 1018 and 1133 $cm^{-1}$. c) Normal modes of the [2] series characterized by C-N shrinking of the donor termination, stretching of the C-C central bond of the polyyne chain and benzonitrile ring deformation. These modes are centered around 930 $cm^{-1}$. d) Alternated stretching and shrinking of the single bonds of the polyyne bridge, coupled with benzonitrile ring deformation and C-H bending. These vibrational bands are located around 1500 $cm^{-1}$. e) Bending of the C-H bonds of the donor ring of the C[2] and D[2] molecules, centered at 1318 and 1331 $cm^{-1}$, respectively. f) Most intense vibrational modes of ynamines with [2] triple bonds in the region from 1350 to 1450 $cm^{-1}$. All these modes have in common the shrinking of the central C-C bond and the stretching of the single bonds at the borders of their sp chain.

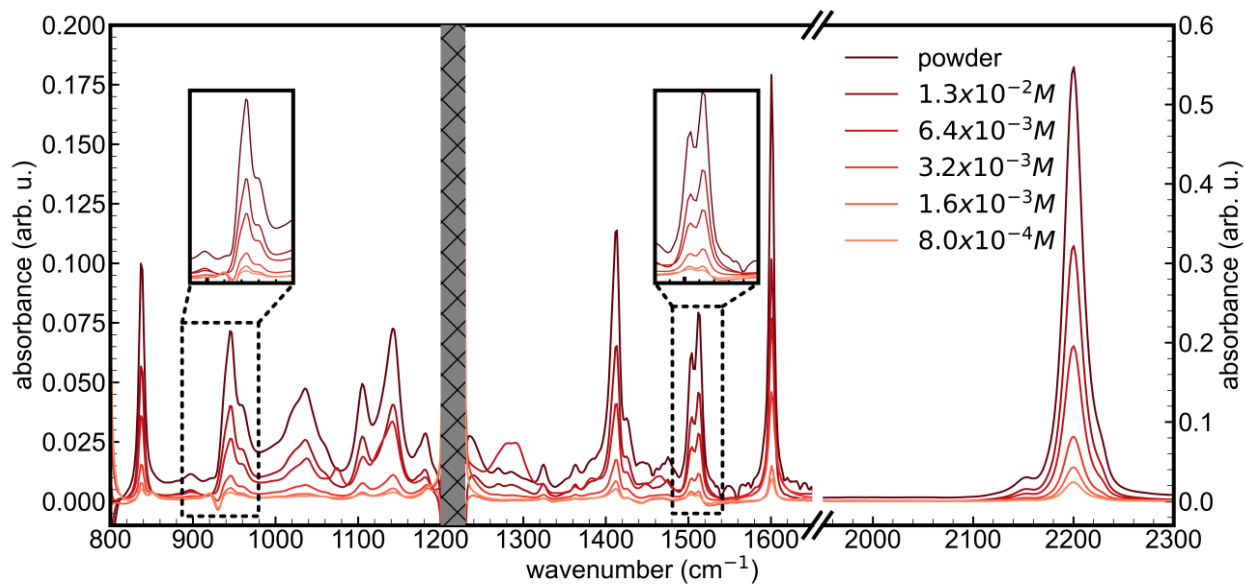

***Figure S6*** *Comparison of the IR spectra of molecule A[2] in both solid-state and solution. The darker curve reports the powder spectra, while the other lines show decreasing concentrations (displayed in the legend) IR spectra of the same molecule. The insets focus on the structure of the bands at around 950 and 1500 cm$^{-1}$. The grayed region covers solvent (chloroform) strong signals.*

## Analysis of DFT calculations performed on a single molecule, dimer and crystal

To assess the relevance of solid-state effects and intermolecular packing on the vibrational response of these materials, periodic boundary conditions DFT calculations have been carried out for the case of A[3] crystal, using the CRYSTAL17 code[1]. Starting from the experimentally derived XRD structure [2], a full-geometry optimization of both cell parameters and atomic coordinates has been carried out. To this aim the functional PBE0 has been used together with a revised pob-DVZP basis set [3], introducing Grimme's D2 correction for dispersion interactions with the parameters proposed in Quarti et al. [4]. The calculation has been repeated also for the isolated molecule and the dimer, whose starting geometry can be derived considering closely packed dimers occurring in the crystal, by using the same setup, to carry out a correct comparison with the spectra computed for the crystal.

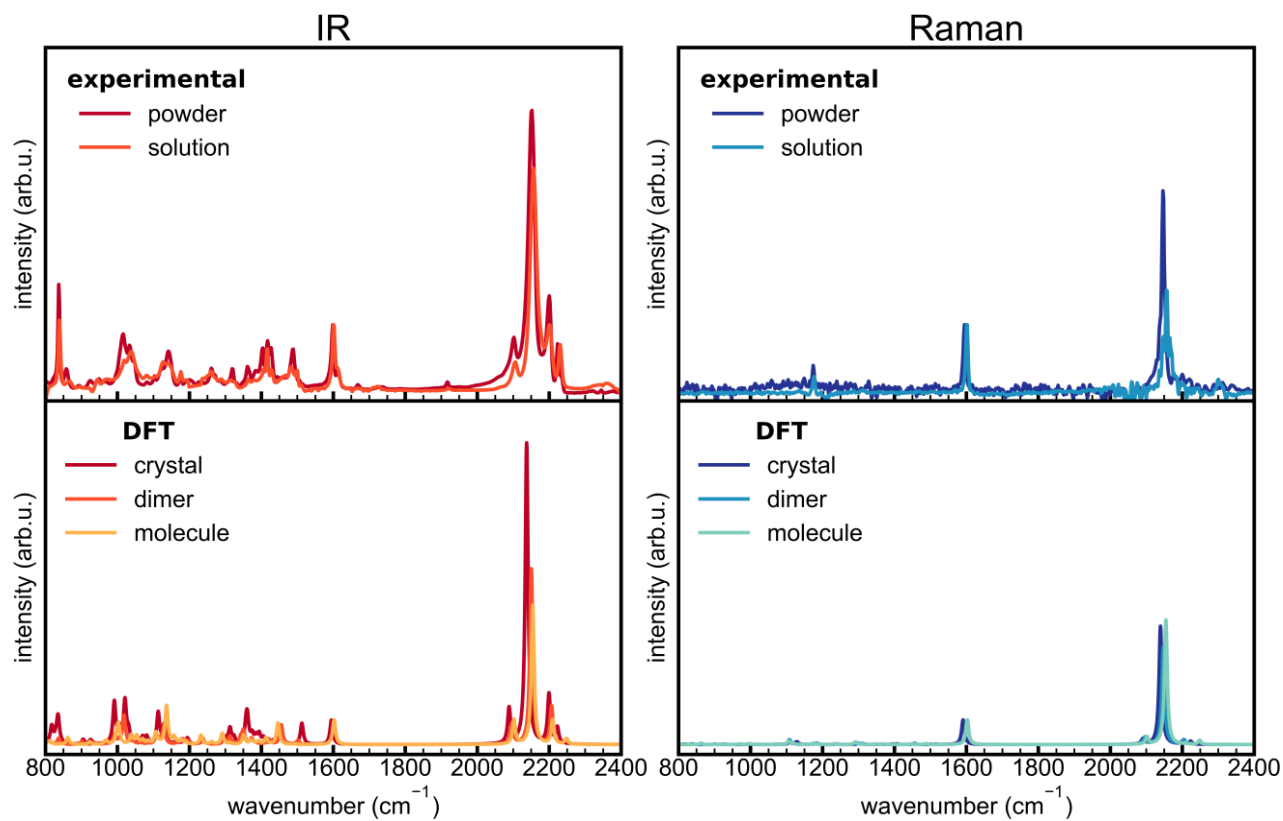

***Figure S7*** *(Upper panels) Relationship between experimental spectra of powder and solution of the A[3] molecule, probed with IR and Raman spectroscopy. (Lower panels) Comparison between DFT calculations performed on a single molecule, dimer and crystal. The simulated spectra were multiplied by a common factor (0.9399) to match the experimental position of the phenyl-ring breathing mode at 1597 cm$^{-1}$.*

## Energetic and spectral investigation of the molecular conformations of A[4]

We have assessed the possible impact of molecular conformation on the IR spectra of A[4] molecule by DFT calculations at the B3LYP/6-31G(d,p) level, a CPU-effective method that is convenient for such screening tasks. The results reported in Figure S8 show two thermally accessible conformers (b, c) at higher energies than the one reported in the main text (a). The simulated spectra display some conformation dependence for the bands at about 1540 cm$^{-1}$ and 950 cm$^{-1}$. The signal in the 1300 – 1500 cm$^{-1}$ range can be modulated by conformational effects as well.

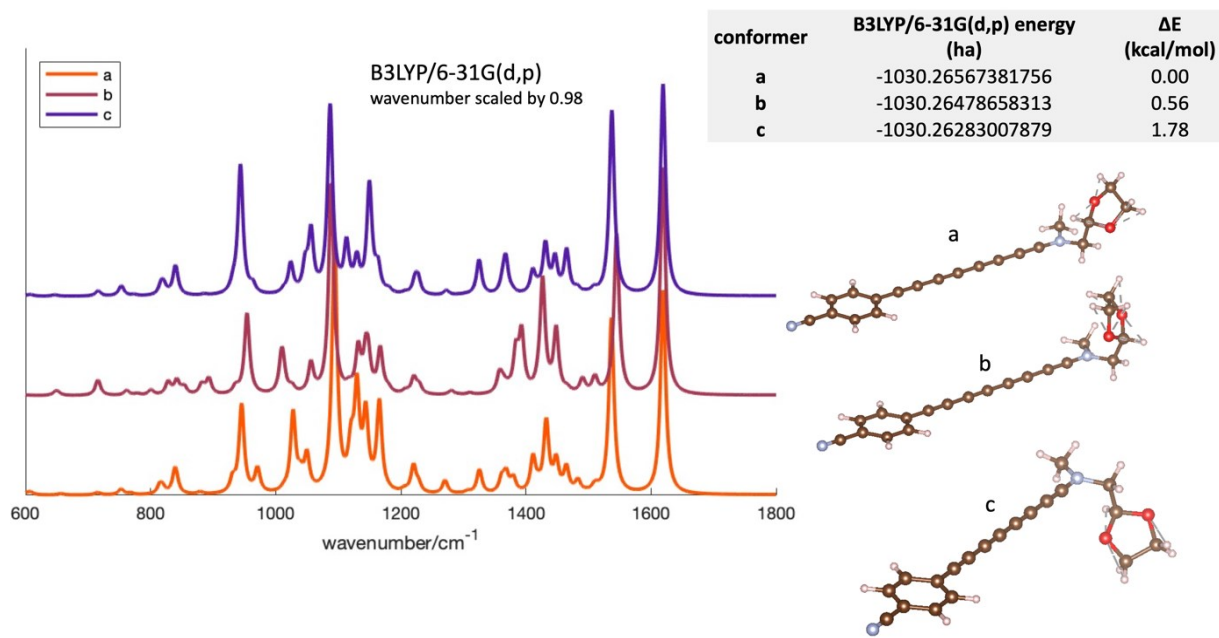

***Figure S8*** *Simulated IR spectra of three low-energy conformations of A[4] molecule obtained by rotations around the dihedral angle around the central CC bond of the donor group. The position of the bands at about 1540 cm$^{-1}$ and 950 cm$^{-1}$ shows conformational dependence. The reported data have been obtained from B3LYP/6-31G(d,p) calculations, and the wavenumber axis has been uniformly scaled by 0.98 (a usual scale factor for this method, adopted in π-conjugated systems to ease comparison with experimental data* [5]*). The three-dimensional representation of the equilibrium geometries of the three conformers is also reported, together with the associated energies.*

| | B3LYP/6-31G(d,p) | | | | | PBE1PBE vs. B3LYP ratio | | | |
|---|---|---|---|---|---|---|---|---|---|
| | A[n] | B[n] | C[n] | D[n] | | A[n] | B[n] | C[n] | D[n] |
| [2] | 9268.62 | 8963.16 | 9213.76 | 8997.59 | [2] | 0.95 | 1.00 | 0.97 | 0.97 |
| [3] | 16408.29 | 16085.07 | 16386.55 | 15658.6 | [3] | 0.94 | 0.98 | 0.95 | 0.96 |
| [4] | 25697.73 | 25361.06 | 25717.18 | 24223.6 | [4] | 0.93 | 0.96 | 0.94 | 0.94 |
| | PBE1PBE/cc-pVTZ | | | | | | | | |
| | A[n] | B[n] | C[n] | D[n] | | | | | |
| [2] | 8799.23 | 8932.46 | 8897.23 | 8725.26 | | | | | |
| [3] | 15402.3 | 15763.14 | 15632.62 | 14998.29 | | | | | |
| [4] | 23786.01 | 24425.07 | 24177.56 | 22868.36 | | | | | |
| | RHF/cc-pVTZ | | | | | PBE1PBE vs HF ratio | | | |
| | A[n] | B[n] | C[n] | D[n] | | A[n] | B[n] | C[n] | D[n] |
| [2] | 2567.24 | 2682.76 | 2973.9 | 2812.48 | [2] | 3.43 | 3.33 | 2.99 | 3.10 |
| [3] | 3321.59 | 3465.18 | 3814.78 | 3613.82 | [3] | 4.64 | 4.55 | 4.10 | 4.15 |
| [4] | 3962.86 | 4055.7 | 4578.65 | 4143.46 | [4] | 6.00 | 6.02 | 5.28 | 5.52 |

**Table S3** *Vector component of the first electronic hyperpolarizabilities ($\beta^e$) of push-pull polyynes A[n], B[n], C[n] and D[n], obtained by means of Quantum Chemical calculations at different levels of theory. Values are in atomic units (8.641 X $10^{-33}$ esu).*

| | B3LYP/6-31G(d,p) | | | | | PBE1PBE vs. B3LYP ratio | | | |
|---|---|---|---|---|---|---|---|---|---|
| | A[n] | B[n] | C[n] | D[n] | | A[n] | B[n] | C[n] | D[n] |
| [2] | 7671.96 | 8170.34 | 7302.58 | 6591.59 | [2] | 0.95 | 0.99 | 0.97 | 1.01 |
| [3] | 13128.47 | 14557.5 | 12809.36 | 11831.22 | [3] | 0.98 | 0.96 | 0.99 | 1.00 |
| [4] | 21361.3 | 22923.32 | 20666.33 | 18875.96 | [4] | 0.94 | 0.94 | 0.98 | 0.98 |
| | PBE1PBE/cc-pVTZ | | | | | | | | |
| | A[n] | B[n] | C[n] | D[n] | | | | | |
| [2] | 7311.76 | 8078.06 | 7111.19 | 6646.89 | | | | | |
| [3] | 12883.13 | 13996.73 | 12736.9 | 11834.88 | | | | | |
| [4] | 20182.99 | 21568.93 | 20155.89 | 18561.38 | | | | | |
| | RHF/cc-pVTZ | | | | | PBE1PBE vs HF ratio | | | |
| | A[n] | B[n] | C[n] | D[n] | | A[n] | B[n] | C[n] | D[n] |
| [2] | 4075.49 | 4121.76 | 4361.03 | 4086.04 | [2] | 1.79 | 1.96 | 1.63 | 1.63 |
| [3] | 5414.43 | 5543.45 | 5432.27 | 4928.21 | [3] | 2.38 | 2.52 | 2.34 | 2.40 |
| [4] | 6452.17 | 6783.17 | 6740.31 | 5757.88 | [4] | 3.13 | 3.18 | 2.99 | 3.22 |

**Table S4** *Vector component of the first vibrational hyperpolarizabilities ($\beta^v$) of push-pull polyynes A[n], B[n], C[n] and D[n], obtained by means of Quantum Chemical calculations at different levels of theory. Values are in atomic units (8.641 X $10^{-33}$ esu).*

# Synthesis and characterization of new compounds

**General information.** Commercially available morpholine (99%, Sigma Aldrich), $Na_3PO_4$ (96%, Sigma Aldrich) and $Al_2O_3$ (Brockmann Grade I, basic, Alfa Aesar) were used without further purification. All solvents for chromatography were used without further purification. Acetonitrile for reaction was distilled over $P_2O_5$.

$^1H$ and $^{13}C$ NMR spectra were recorded on a Bruker Avance 500 MHz using an inverse broadband (BBI) probehead. For all the $^1H$ NMR spectra, the chemical shifts are given in ppm relative to the solvent residual peaks ($CDCl_3$, $^1H$ NMR: 7.26 ppm, $^{13}C$ NMR: 77.16 ppm). Coupling constants are given in Hz.
HRMS spectra were recorded using a Bruker apex ultra FT-ICR spectrometer with ESI ion source or a JEOL MALDI-TOF JMS-S300 Spiral-TOF-plus spectrometer. DCTB (*trans*-2-[3-(4-*tert*-butylphenyl)-2-methyl-2-propenylidene]malononitrile) was used as a matrix for MALDI measurements.

**Compound D[2]:**

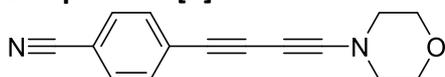

**$^1$H NMR** (500 MHz, $CDCl_3$): δ = 7.56-7.54 (m, 2H), 7.47-7.44 (m, 2H), 3.74-3.71 (m, 4H), 3.22-3.18 (m, 4H). **$^{13}$C NMR** (126 MHz, $CDCl_3$): δ = 132.0, 132.0, 128.4, 118.6, 110.9, 89.9, 80.0, 78.5, 65.9, 51.1, 51.0. **HRMS(ESI)**: *m/z* calculated for $C_{15}H_{12}N_2ONa$: 259.08418 [M+Na$^+$]; found: 259.08417.

**Compound D[3]:**

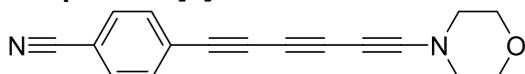

**$^1$H NMR** (500 MHz, $CDCl_3$): δ = 7.61-7.57 (m, 2H), 7.56-7.52 (m, 2H), 3.75-3.70 (m, 4H), 3.24-3.20 (m, 4H). **$^{13}$C NMR** (126 MHz, $CDCl_3$): δ = 133.1, 132.2, 127.2, 118.5, 112.1, 85.7, 79.8, 76.4, 70.9, 67.0, 66.0, 53.6, 51.0. **HRMS(ESI)**: *m/z* calculated for $C_{17}H_{12}N_2ONa$: 283.0842 [M+Na$^+$]; found: 283.0809.

**Compound D[4]:**

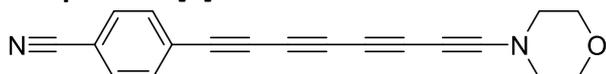

**$^1$H NMR** (500 MHz, $CDCl_3$): δ = 7.62-7.59 (m, 2H), 7.58-7.56 (m, 2H), 3.74-3.70 (m, 4H), 3.25-3.21 (m, 4H). **$^{13}$C NMR** (126 MHz, $CDCl_3$): δ = 133.5, 132.2, 126.5, 118.3, 112.7, 84.3, 79.2, 75.5, 70.8, 68.0, 67.2, 66.0, 63.4, 54.8, 50.9. **HRMS(MALDI-TOF)**: *m/z* calculated for $C_{19}H_{12}N_2O$: 284.0944 [M$^+$]; found: 284.0950.

# NMR Spectra

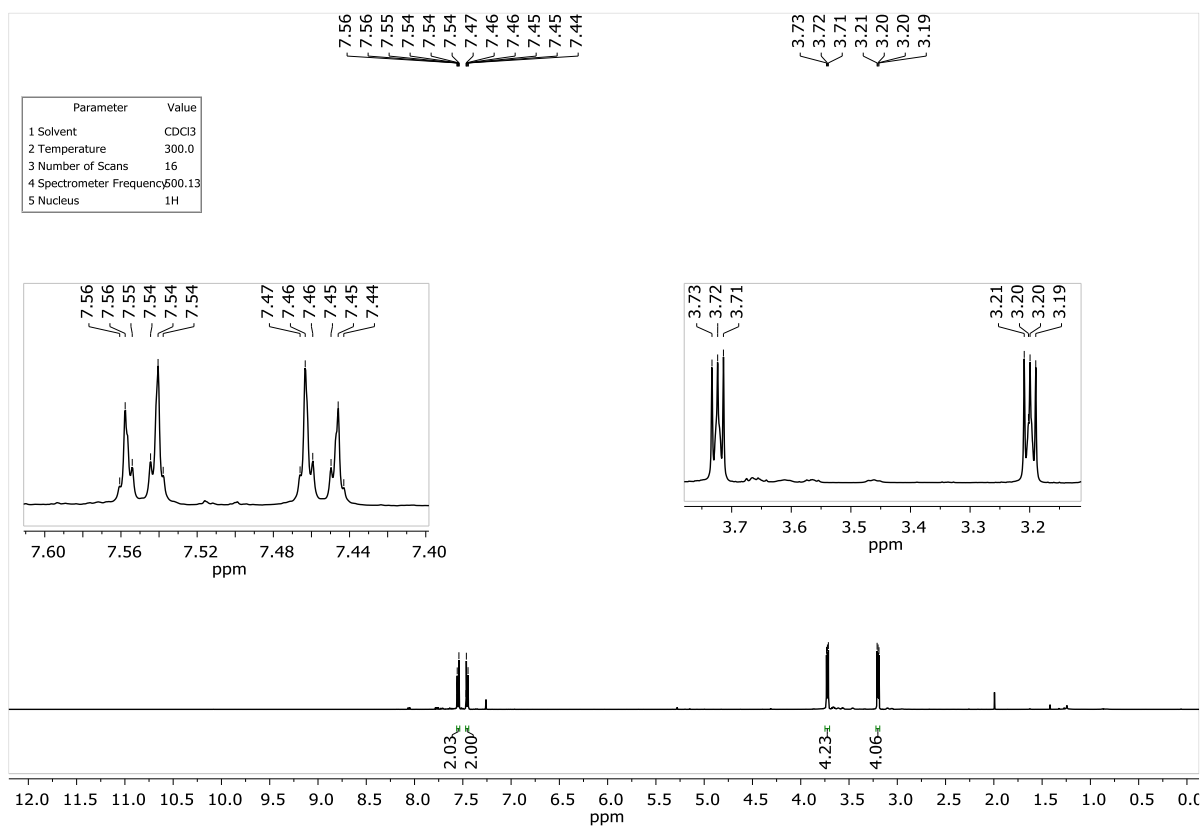

***Figure S9*** *$^1$H NMR spectrum of D[2] (500 MHz, CDCl$_3$, 300 K).*

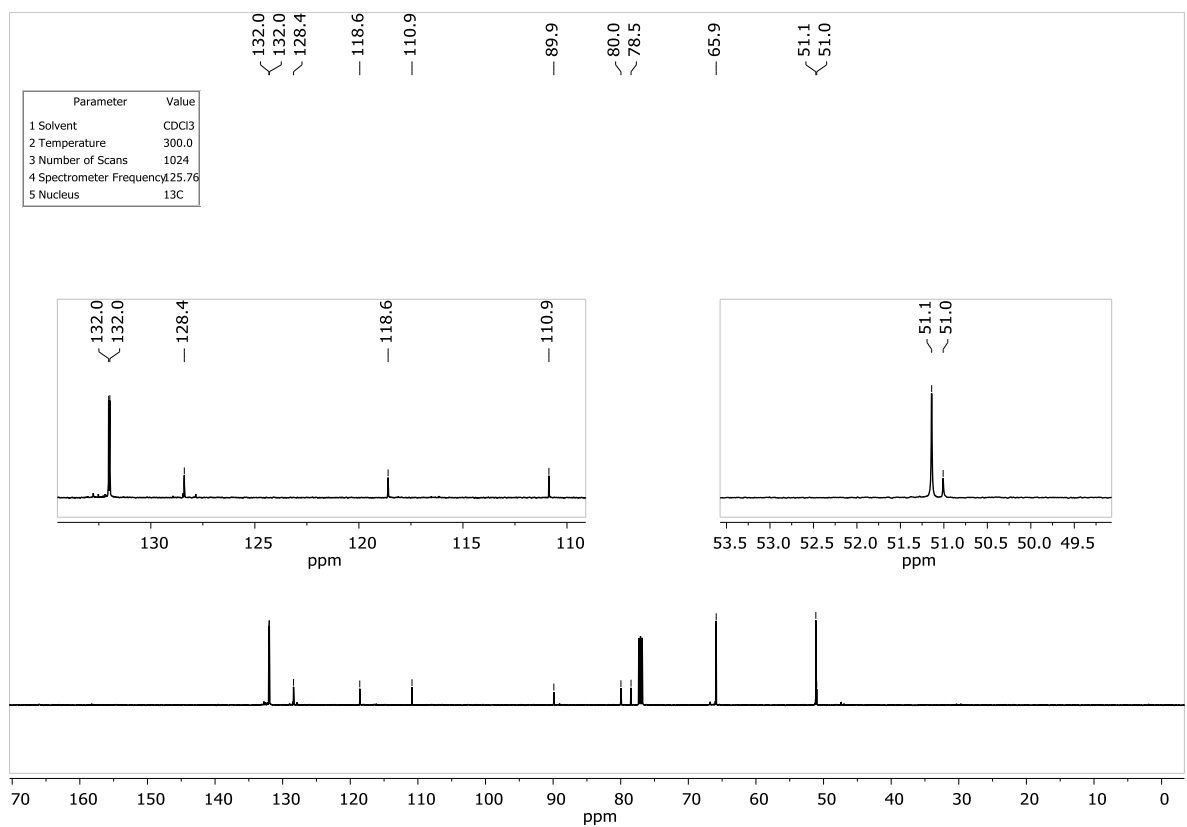

***Figure S10*** *$^{13}$C NMR spectrum of D[2] (126 MHz, CDCl$_3$, 300 K).*

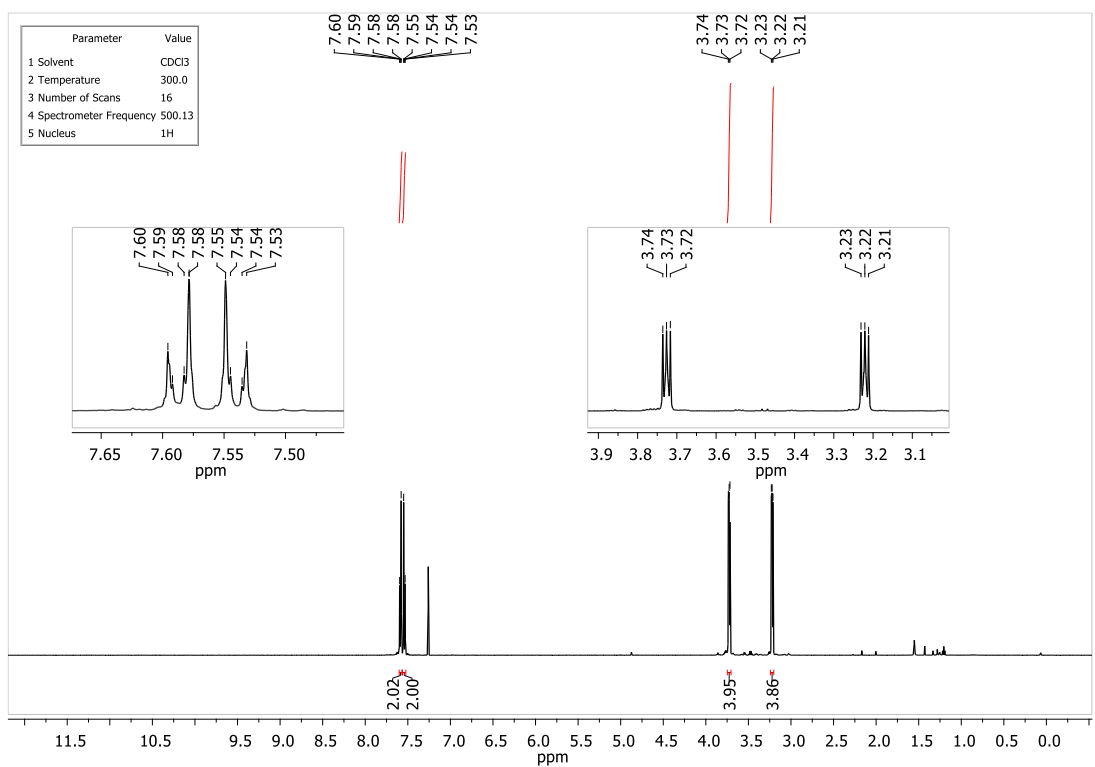

***Figure S11*** $^1$H NMR spectrum of D[3] (500 MHz, CDCl$_3$, 300 K).

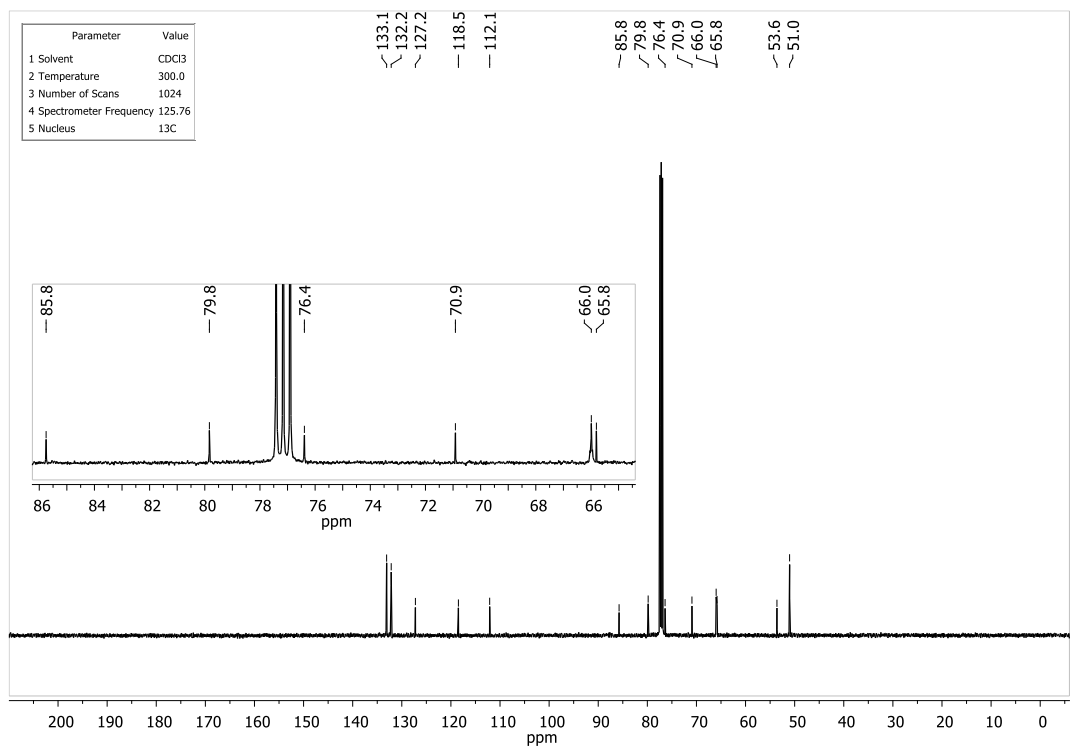

***Figure S12*** $^{13}$C NMR spectrum of D[3] (126 MHz, CDCl$_3$, 300 K).

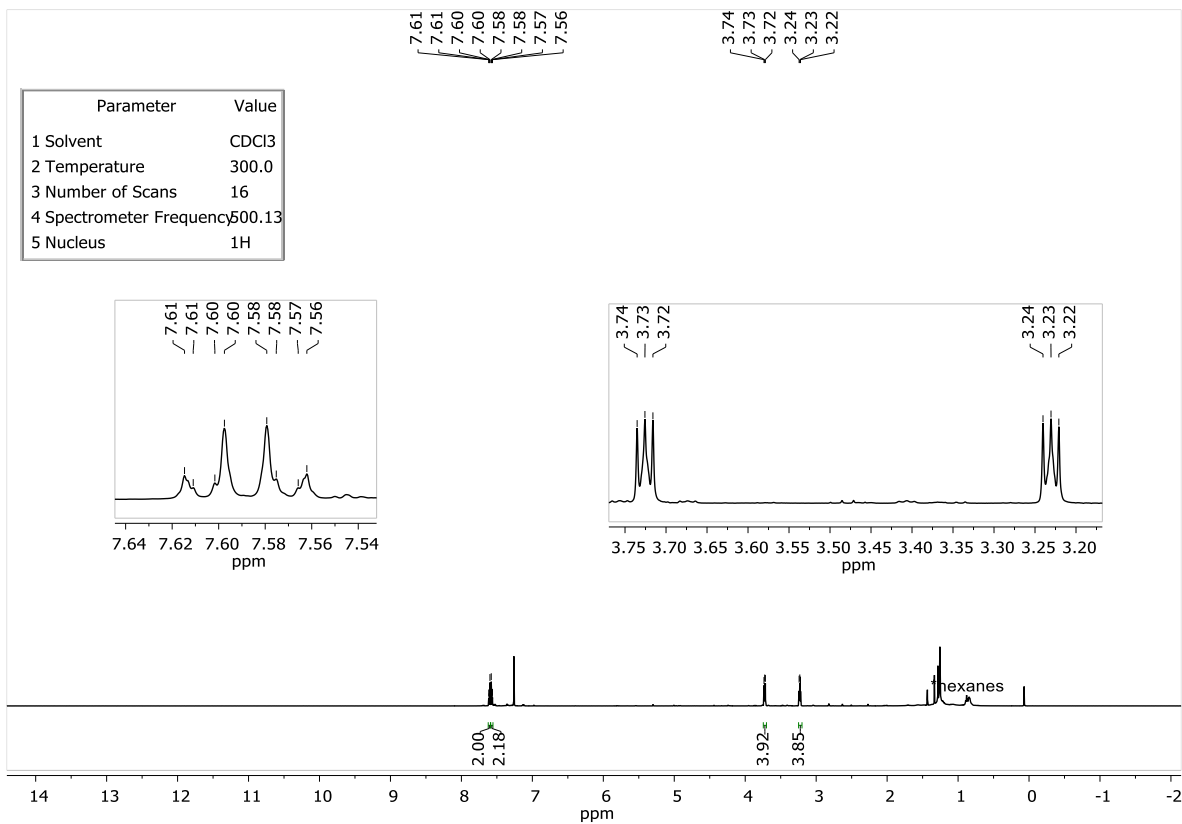

*Figure S13* [1]H NMR spectrum of D[4] (500 MHz, CDCl$_3$, 300 K).

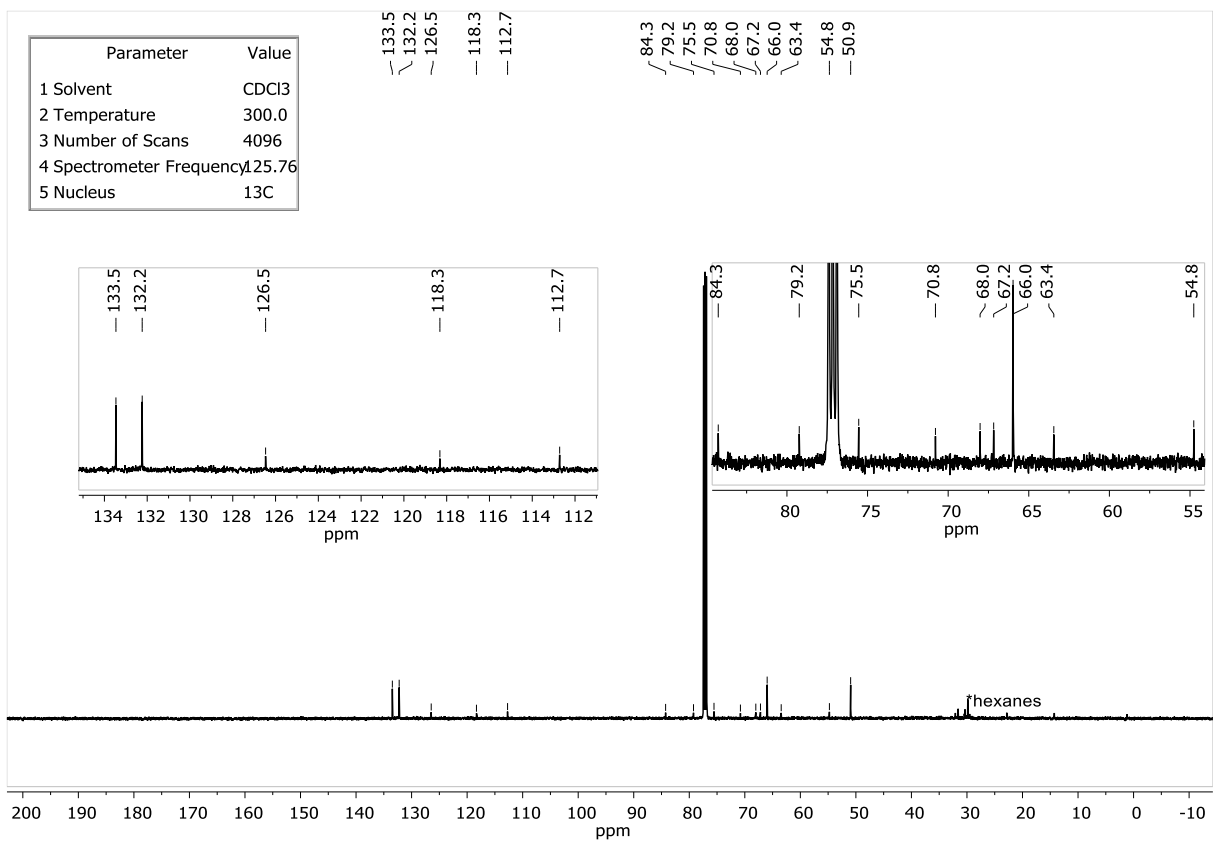

*Figure S14* [13]C NMR spectrum of D[4] (126 MHz, CDCl$_3$, 300 K).

## Single crystal X-ray diffraction

X-ray diffraction data were collected with Rigaku Synergy diffractometer (ω scan technique). The space groups were determined from systematic absences and subsequent least-squares refinement. Lorentz and polarization corrections were applied. The structure was solved by direct methods and refined by full-matrix, least-squares on $F^2$ by use of the SHELX package[6] and Olex2 software[7]. Hydrogen atom positions were calculated and added to the structure factor calculations but were not refined.

CCDC 2087228 contains the supplementary crystallographic data for this paper. These data can be obtained free of charge via www.ccdc.cam.ac.uk/data_request/cif, or by emailing data_request@ccdc.cam.ac.uk, or by contacting The Cambridge Crystallographic Data Centre, 12 Union Road, Cambridge CB2 1EZ, UK; fax: + 44 1223 336033.

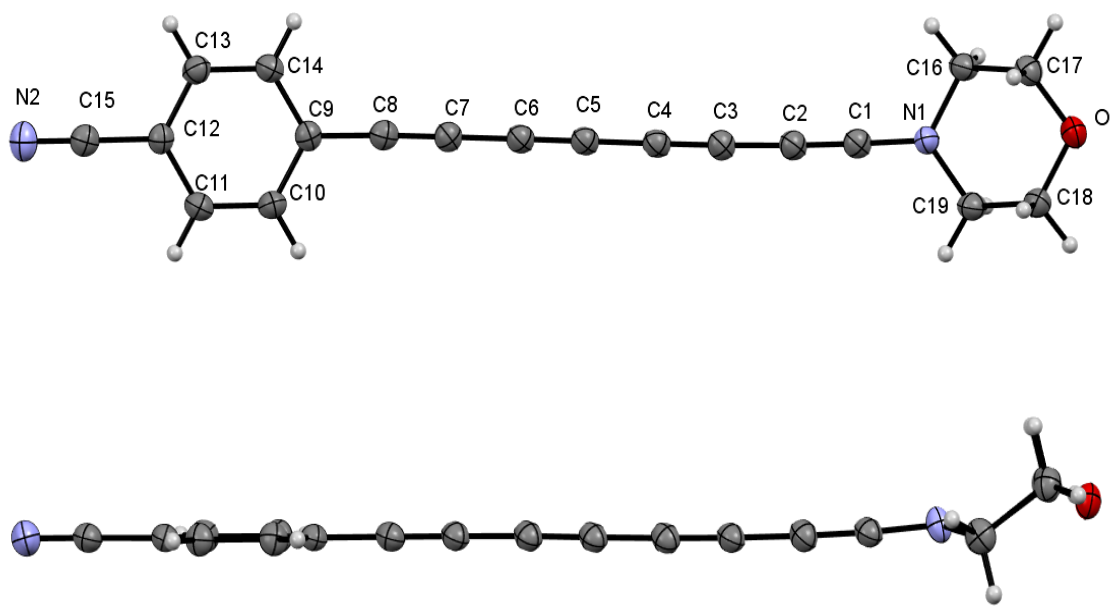

**Figure S15** *Molecular structure of compound D[4]. The thermal ellipsoids are set at a 50% probability level (CCDC 2087228).*

| Compound | D[4] (CCDC 2087228) |
|---|---|
| *Crystal data* | |
| Chemical formula | $C_{19}H_{12}N_2O$ |
| $M_r$ | 284.31 |
| Crystal system, space group | Monoclinic, $P2_1/n$ |
| Temperature (K) | 100 |
| $a, b, c$ (Å) | 7.022 (3), 7.846 (3), 27.056 (8) |
| $β$ (°) | 91.99 (3) |
| $V$ (Å$^3$) | 1489.6 (10) |
| $Z$ | 4 |
| Radiation type | Cu $K\alpha$ |
| $\mu$ (mm$^{-1}$) | 0.64 |
| Crystal size (mm) | 0.25 × 0.16 × 0.05 |
| *Data collection* | |
| Diffractometer | XtaLAB Synergy R, DW system, HyPix-Arc 150 |
| Absorption correction | - |
| No. of measured, independent and observed reflections | 11678, 3058, 2599 {$I > 2σ(I)$} |
| $R_{int}$ | 0.021 |
| $(\sin ϑ/λ)_{max}$ (Å$^{-1}$) | 0.629 |
| *Refinement* | |
| $R[F^2 > 2σ(F^2)]$, $wR(F^2)$, $S$ | 0.039, 0.112, 1.04 |
| No. of reflections | 3058 |
| No. of parameters | 199 |
| H-atom treatment | H-atom parameters constrained |
| $Δρ_{max}, Δρ_{min}$ (e Å$^{-3}$) | 0.17, −0.23 |

**Table S3** X-ray crystallography details of D[4].